\documentclass[fleqn]{aa}
\usepackage{times}
\usepackage{graphics}

\begin{document}


   \title{Line formation in solar granulation}

   \subtitle{IV. [O\,{\sc i}], O\,{\sc i} and OH lines and the photospheric O abundance}

   \author{Martin Asplund$^{1}$,
           Nicolas Grevesse$^{2,3}$, A. Jacques Sauval$^{4}$,
           Carlos Allende Prieto$^{5}$ and
	   Dan Kiselman$^{6}$
          }

   \institute{
$^{1}$ Research School of Astronomy and Astrophysics,
Mt. Stromlo Observatory, Cotter Rd., Weston,
ACT 2611, Australia\\
$^{2}$ 
Centre Spatial de Li\`ege, Universit\'e de Li\`ege,
avenue Pr\'e Aily, B-4031 Angleur-Li\`ege, Belgium \\
$^{3}$ Institut d'Astrophysique et de G\'eophysique, 
Universit\'e de Li\`ege, All\'ee du
6 ao\^ut, 17, B5C, B-4000 Li\`ege, Belgium \\
$^{4}$ Observatoire Royal de Belgique, avenue circulaire, 3,
B-1180 Bruxelles, Belgium \\
$^{5}$ McDonald Observatory and Department of Astronomy,  
University of Texas, Austin, TX 78712-1083, USA \\
$^{6}$ The Institute for Solar Physics of the Royal Swedish Academy of Sciences, 
AlbaNova University Centre, SE-106 91 Stockholm, Sweden
             }

   \offprints{\email{martin@mso.anu.edu.au}}

   \date{{\em Accepted for Astronomy \& Astrophysics}}

\abstract{
The solar photospheric 
oxygen abundance has been determined from [O\,{\sc i}], O\,{\sc i},
OH vibration-rotation and OH pure rotation lines by means of
a realistic time-dependent, 3D, hydrodynamical model of the solar atmosphere.
In the case of the O\,{\sc i} lines, 3D non-LTE calculations have been
performed, revealing significant departures from LTE as a result of
photon losses in the lines. 
We derive a solar oxygen abundance of log\,$\epsilon_{\rm O} = 8.66 \pm 0.05$.
All oxygen diagnostics yield highly consistent
abundances, in sharp contrast with the results of classical 1D model atmospheres.
This low value is in good agreement with
measurements of the local interstellar medium and nearby B stars.
This low abundance is also supported by the excellent correspondence
between lines of very different line formation sensitivities, and between
the observed and predicted line shapes and center-to-limb variations.
Together with the corresponding down-ward revisions of the solar carbon, nitrogen
and neon abundances, the resulting significant decrease in solar metal mass fraction
to $Z = 0.0126$
can, however, potentially spoil the impressive agreement between
predicted and observed sound speed in the solar interior
determined from helioseismology.
}

\authorrunning{M. Asplund et al.}
\titlerunning{Solar line formation: IV. The photospheric O abundance}

   \maketitle

\section{Introduction}

Oxygen is the most abundant element in the Universe with
a non-Big Bang nucleosynthesis origin.
As a consequence, oxygen plays a central role in many different fields of
astrophysics ranging from supernova physics and galaxy evolution to dating stars and
production of the light elements through cosmic ray spallation.
Yet it appears that in many crucial objects for which accurate knowledge of
the oxygen abundances is necessary
the oxygen content is hotly debated. Recent disputes revolve around the
overabundance of oxygen in metal-poor halo stars
(see Asplund \& Garc\'{\i}a P{\'e}rez 2001; Nissen et al. 2002
and references therein), the Galactic radial abundance gradient
(Rolleston et al. 2000; Cunha \& Daflon 2003), 
and, astonishingly, the solar oxygen abundance.
Partly these disagreements stem from differences in the adopted input data
(e.g. $gf$-values, effective temperatures $T_{\rm eff}$, surface gravities log\,$g$)
but more importantly they reflect
the choice of spectral lines to derive the abundances using classical
1D stellar model atmospheres.
In particular in the solar case, the freedom of parameters to obtain
consistency is very restricted yet the discrepancy is present in full.

Until recently the commonly adopted solar oxygen abundance was
log\,$\epsilon_{\rm O} = 8.93 \pm 0.04$\footnote{
On the customary abundance scale defined as
$\epsilon {\rm (X)} = 10^{12} \cdot N{\rm (X)}/ N{\rm (H)}$}
(Anders \& Grevesse 1989).
This historically 
high abundance was suggested by analyses of the forbidden [O\,{\sc i}] 630.0\,nm
line (Lambert 1978) as well as OH vibration-rotation and pure rotation
lines in the infrared (Grevesse et al. 1984; Sauval et al. 1984) 
using the 1D hydrostatic Holweger-M\"uller
(1974) semi-empirical model of the solar atmosphere and LTE line formation.
On the other hand, a much lower abundance is indicated by the permitted
high-excitation O\,{\sc i} lines, most noteworthy the
IR triplet at 777\,nm, when employing the same model atmosphere with non-LTE
line formation.
This discrepancy of about 0.2\,dex between different abundance
indicators have often been blamed on over-estimated
departures from local thermodynamic equilibrium (LTE) for the
O\,{\sc i} lines (Tomkin et al. 1992; Takeda 1994).
Indeed, the LTE abundance of the triplet is close to
the [O\,{\sc i}] and OH-based abundance with the Holweger-M\"uller model.
Such argumentation, however, ignores the available observational evidence
such as the center-to-limb variation that the triplet is definitely
not formed in LTE (Altrock 1968; Sedlmayr 1974; 
Kiselman 1993; Kiselman \& Nordlund 1995).
Uncertainties surround not only the O\,{\sc i} lines but also the
other indicators. The [O\,{\sc i}] line is very weak and 
blended with a Ni\,{\sc i} line
(Lambert 1978; Reetz 1998; Allende Prieto et al. 2001).
The OH lines are, like all molecules in the relevant temperature regime
of the solar atmosphere, very temperature sensitive and therefore
susceptible to surface inhomogeneities like granulation
(Kiselman \& Nordlund 1995; Asplund \& Garc\'{\i}a P{\'e}rez 2001), which
are ignored in classical 1D model atmospheres.
Unfortunately, no guidance regarding the solar oxygen abundance is
available from meteorites, contrary to the situation for most other elements
(Grevesse \& Sauval 1998). Like carbon and nitrogen, oxygen is a highly volatile
element which has partly escaped from the most primitive meteorites,
the CI-chondrites, leaving only a fraction of the original amount
(10\%, 3\% and 54\% of C, N and O, respectively).

Recently, Allende Prieto et al. (2001) suggested a significant down-ward
revision of the solar oxygen abundance to
log\,$\epsilon_{\rm O} = 8.69 \pm 0.05$ based on a study of the
[O\,{\sc i}] 630.0\,nm line using a time-dependent 3D hydrodynamical
model of the solar atmosphere (Asplund et al. 2000b). Most
of the 0.24\,dex difference with the previous higher value is
attributed to a previously ignored 
Ni\,{\sc i} blend ($-0.13$\,dex), although the
difference between 3D and 1D models compounds the effect ($-0.08$\,dex).
In addition, a revision of the transition probability of the line causes a further
0.03\,dex decrease. 
A similar low value was obtained by Holweger (2001) driven
primarily by permitted O\,{\sc i} lines: 
log\,$\epsilon_{\rm O} = 8.74 \pm 0.08$.
This new low O abundance largely resolves
another puzzling problem with the old high oxygen abundance, namely that
the Sun is apparently oxygen-rich compared with the present-day local interstellar
medium (e.g. Meyer et al. 1998; Andr\'e et al. 2003) and nearby B stars
(e.g. Cunha \& Lambert 1994; Kilian et al. 1994;
Sofia \& Meyer 2001). Since the birth of the Sun
some 4.5\,Gyr ago, the gas in the solar neighborhood from which these
hot stars have recently formed should have been further chemically enriched
by nuclear-processed ejecta from dying stars.
Several explanations like migration of the solar Galactic orbit
have been proposed to address this conundrum but
they may all be superfluous if the new low solar oxygen abundance is confirmed.

Relying solely on a single line for an as important quantity as the
solar O abundance is no doubt highly unsatisfactory, in particular given
the significant Ni blend of the [O\,{\sc i}] 630.0\,nm line.
It is therefore clearly of importance to extend the work by
Allende Prieto et al. (2001) to include also the O\,{\sc i} and OH
lines analysed with the same 3D hydrodynamical model atmosphere.
Such a study is presented here, which firmly establishes the reality
of the low solar photospheric oxygen abundance.
It is noteworthy that for the first time ever the different line indicators give
consistent results, which strongly supports the high degree of
realism obtained with the new generation of 3D model atmospheres
suggested by previous studies.

\section{Analysis}

\subsection{Atomic and molecular data}

\noindent
{\bf [O\,{\sc i}] lines:}
The $gf$-value of the [O\,{\sc i}] 630.03\,nm line has
recently been revised to log\,$gf = -9.717$
(Storey \& Zeippen 2000), which we adopt here. This is a slight revision
from the previously commonly adopted value of log\,$gf = -9.75$ (Lambert 1978).
The 630.03\,nm feature has a non-negligible contribution from a Ni\,{\sc i} line
(Lambert 1968; Reetz 1998; Allende Prieto et al. 2001) which must be taken
into account. Johansson et al. (2003) has very recently measured the
transition probability of the Ni line and found log\,$gf = -2.11$.
However, since only the product $\epsilon_{\rm Ni} \cdot gf$ enters
the $\chi^2$-analysis of the [O\,{\sc i}] 630.0\,nm line profile fitting
as a free parameter (Allende Prieto et al. 2001), this new value will not modify the
derived O abundance but only the solar Ni abundance suggested by this one line.
The isotopic splitting for the Ni line
also determined by Johansson et al. (2003)
is taken into account for the synthesis of the 630.03\,nm line.
The corresponding data for the weaker [O\,{\sc i}] 636.37\,nm line
is log\,$gf = -10.185$ (Storey \& Zeippen 2000). In the analysis of
this transition care must be exercised
due to problematic blending of CN lines and a Ca\,{\sc i} auto-ionization
line at the wavelength of the [O\,{\sc i}] 636.37\,nm line.

\noindent
{\bf O\,{\sc i} lines:} As described below, we chose to retain only
six permitted O\,{\sc i} lines (615.81, 777.19, 777.41, 777.53,
844.67 and 926.6\,nm) for the abundance analysis to minimise
the errors stemming from blends, uncertain $gf$-values,
continuum placement and too weak spectral lines.
The transition probabilities and lower level excitation potentials
for the five lines were taken from the NIST database\footnote{
http://physics.nist.gov/cgi-bin/AtData/main\_asd}.
The VALD database\footnote{
http://www.astro.uu.se/$\sim$vald}
(Piskunov et al. 1995; Kupka et al. 1999) was
consulted for radiative broadening and central wavelengths.
Collisional broadening by H (commonly referred to as van der Waals broadening)
was computed from the quantum mechanical theory of
Anstee \& O'Mara (1995), Barklem \& O'Mara (1997) and Barklem et al. (1998).
This removes the need for the conventional enhancement factors to the
Uns\"old (1955) classical recipe for collisional broadening.
Some of the relevant line properties are listed in Table \ref{t:nlte}.


\noindent
{\bf OH vibration-rotation and pure rotation lines:}
We selected the best 70 vibration-rotation lines of the (1,0), (2,1) and (3,2) bands
as well as the best 127 pure rotation lines belonging to the 0-0, 1-1, 2-2 and 3-3 
bands in the infrared solar spectrum.  
The dissociation energy of OH is very accurately known since quite a long time:
$D_0$ (OH)=4.392 $\pm$ 0.005 eV (Carlone \& Dalby 1969).
The molecular partition functions and equilibrium constants have been taken
from Sauval \& Tatum (1984).  
Lower level excitation potentials were adopted from M\'elen et al. (1995).
The {\it gf}-values were calculated by E. van Dishoeck (private communication)
adopting the Electric Dipole Moment Function (EDMF) of Nelson et al. (1990). 
We checked that this EDMF leads to more consistent results than other calculated EDMFs.
We note that this EDMF has also been selected by Goldman et al. (1998)
in their recent data base for the OH X $^2\Pi$ ground state. A direct comparison
of the {\it gf}-values with those of Goldman et al. (1998) reveal essentially
identical results.

\subsection{Observational data}

For the analysis of the forbidden [O\,{\sc i}] and permitted
O\,{\sc i} lines which are all located in the optical region
(600-850\,nm) the solar flux atlas of Kurucz et al. (1984) has been employed.
This atlas has a signal-to-noise ratio in the optical of 200-3000, and a
resolving power of about 500,000.
Minor adjustments to the tabulated continuum level were made
for the small wavelength regions directly surrounding the lines.
For the 3D analysis, the oxygen abundances suggested by the individual
lines were determined by profile fitting which is possible in view
of the in general excellent agreement between predicted and observed
line profiles in 3D (e.g. Asplund et al. 2000a,b,c).
This is also the main reason why we employ flux profiles 
rather than the computationally less demanding disk-center intensity
profiles for the oxygen lines: the agreement between predicted
and observed profiles is slightly worse in intensity for the  
O\,{\sc i} lines. Their very significant departures from LTE
modify the intensity profiles more than the corresponding flux profiles.
We emphasize, however, that very similar oxygen abundances
are obtained for the O\,{\sc i}  and [O\,{\sc i}] lines with
flux and disk-center intensity profiles.

The OH vibration-rotation and pure rotation lines are located in
the infrared. Our analysis has made use of the Spacelab-3 
Atlas-3 {\sc atmos}\footnote{
http://remus.jpl.nasa.gov/atmos/ftp.at3.sun.html}
solar disk-center intensity IR observations recorded in November 1994
which are of better quality (i.e. less telluric absorption) than in 
the IR atlas (Farmer \& Norton 1989; Farmer 1994).
The relevant OH lines were identified and their equivalent widths
measured (in milli-Kayser $\equiv$ milli cm$^{-1}$, subsequently converted
to the corresponding values in wavelength units [pm]).

Solar observations of the center-to-limb variation of several spectral lines,
were carried out in October 22, 1997, with the Gregory Coud\'e 
Telescope (GCT) and its Czerny-Turner echelle 
spectrograph (Kneer et al. 1987)
at the Observatorio del Teide (Tenerife, Spain). 
The observations of the O\,{\sc i} 777\,nm triplet have an estimated 
resolving power $R = 86,000$.
Scattered light was
estimated to amount to about $6$\% and corrected
by comparing the observations at the center of the disk with the 
Fourier Transform Spectrograph atlas by Brault \& Neckel (1987).
The observations were done for disk-positions 
$\mu = 1.00, 0.97, 0.87, 0.71, 0.50$ and $0.17$.
The observed center-to-limb variations of the equivalent widths for the O\,{\sc i}
777\,nm triplet are very similar to those measured by M\"uller et al. (1968)
and Altrock (1968) except at the lowest $\mu$-point, 
where our data is less steep.  
More details of the observations and data reduction are provided
in a separate paper (Allende Prieto et al., in preparation).

\subsection{3D hydrodynamical solar model atmosphere}

We employ the same 3D, time-dependent, hydrodynamical model of the solar
atmosphere, which has previously been applied to solar
spectrum line formation for the present series of articles
(Asplund 2000; Asplund et al. 2000b,c; Asplund et al. 2004b,c) 
and elsewhere (Allende Prieto et al. 2001, 2002a, b; Asplund 2004).
This new generation of model atmospheres has proven to be highly
realistic, successfully reproducing a wide range of observational constraints,
such as spectral line shapes, shifts and asymmetries
(Paper I; Allende Prieto et al. 2002a),
helioseismology (Rosenthal et al. 1999; Stein \& Nordlund 2001),
flux distribution and limb-darkening (Asplund et al. 1999b) and granulation
properties (Stein \& Nordlund 1998). It is clear that the new 3D solar
model atmosphere represents a significant improvement
over existing 1D hydrostatic LTE models.

\begin{figure}[t]
\resizebox{\hsize}{!}{\includegraphics{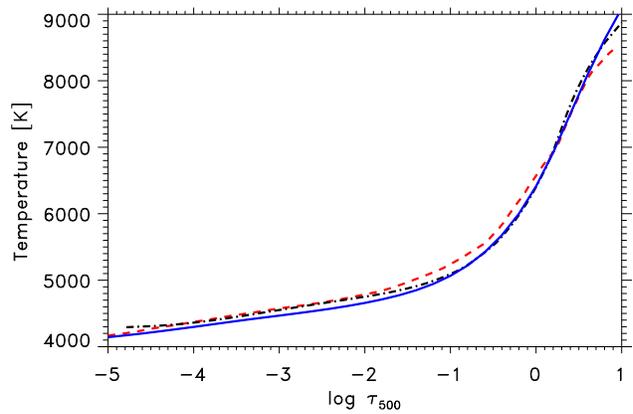}}
\caption{The temporally and spatially averaged temperature structure
of the 3D solar surface convection simulation (solid line) 
used for the 3D spectral line 
formation. The spatial averaging has been performed over
surfaces of the same continuum optical depths at 500\,nm. Note that
the actual 3D simulation extends to much greater optical depths than shown
here. Also shown are the temperature structures for the 1D
Holweger-M\"uller (1974) semi-empirical solar atmosphere (dashed line)
and the 1D {\sc marcs} (Asplund et al. 1997) 
theoretical solar atmosphere (dashed-dotted line).
}
         \label{f:ttau}
\end{figure}

\begin{table*}[t!]
\caption{Temporally and spatially averaged atmospheric stratification of
the 3D solar surface convection simulation used for the 3D spectral line 
formation presented here: temperature, density, gas pressure, electron
pressure and vertical velocity (positive velocities inward)
as well as their rms-scatter. 
The spatial averaging has been performed over
surfaces of the same continuum optical depths at 500\,nm. Note that 
the original 3D model extends significantly above and below the optical
depths for which the mean structure has been interpolated to and tabulated. 
We emphasize that it is not possible to explain all 3D effects by a 
comparison of such an averaged 3D model with existing 1D models due
to the absence of atmospheric inhomogeneities. 
\label{t:model}
}
\begin{tabular}{ccccccccccc}
 \hline
log$\tau_{\rm 500nm}$ & $T$ & $\Delta T_{\rm rms}$ & $\rho$ & $\Delta \rho_{\rm rms}$ &
$P_{\rm gas}$ & $\Delta P_{\rm gas, rms}$ & $P_{\rm e}$ & $\Delta P_{\rm e, rms}$ &
$v_{\rm z}$ & $\Delta v_{\rm z,rms}$ \\ 
& [K] & [K] & [kg/m$^3$] & [kg/m$^3$] & [Pa]   & [Pa] & [Pa] & [Pa] & [km/s] & [km/s] \\
 \hline
 -5.00 &    4143 &    153 &   8.28E-13 &   8.39E-14 &   2.34E+01 &   3.16E+00 &   2.09E-03 &   4.56E-04 &     0.06 &     1.53 \\ 
 -4.80 &    4169 &    161 &   1.09E-12 &   1.16E-13 &   3.10E+01 &   4.40E+00 &   2.73E-03 &   6.39E-04 &     0.04 &     1.46 \\ 
 -4.60 &    4198 &    167 &   1.43E-12 &   1.55E-13 &   4.10E+01 &   5.89E+00 &   3.58E-03 &   8.65E-04 &     0.03 &     1.37 \\ 
 -4.40 &    4230 &    172 &   1.88E-12 &   1.99E-13 &   5.41E+01 &   7.54E+00 &   4.69E-03 &   1.14E-03 &     0.03 &     1.29 \\ 
 -4.20 &    4263 &    176 &   2.47E-12 &   2.44E-13 &   7.13E+01 &   9.30E+00 &   6.11E-03 &   1.47E-03 &     0.03 &     1.22 \\ 
 -4.00 &    4297 &    180 &   3.21E-12 &   2.90E-13 &   9.34E+01 &   1.12E+01 &   7.94E-03 &   1.88E-03 &     0.03 &     1.14 \\ 
 -3.80 &    4332 &    184 &   4.16E-12 &   3.40E-13 &   1.22E+02 &   1.32E+01 &   1.03E-02 &   2.39E-03 &     0.03 &     1.07 \\ 
 -3.60 &    4368 &    188 &   5.36E-12 &   3.99E-13 &   1.58E+02 &   1.55E+01 &   1.32E-02 &   3.03E-03 &     0.03 &     1.02 \\ 
 -3.40 &    4402 &    192 &   6.89E-12 &   4.69E-13 &   2.04E+02 &   1.81E+01 &   1.69E-02 &   3.84E-03 &     0.03 &     0.96 \\ 
 -3.20 &    4435 &    197 &   8.82E-12 &   5.56E-13 &   2.63E+02 &   2.11E+01 &   2.16E-02 &   4.88E-03 &     0.03 &     0.92 \\ 
 -3.00 &    4467 &    202 &   1.13E-11 &   6.75E-13 &   3.38E+02 &   2.46E+01 &   2.74E-02 &   6.22E-03 &     0.04 &     0.89 \\ 
 -2.80 &    4500 &    207 &   1.44E-11 &   8.38E-13 &   4.33E+02 &   2.89E+01 &   3.48E-02 &   7.93E-03 &     0.04 &     0.86 \\ 
 -2.60 &    4535 &    212 &   1.83E-11 &   1.06E-12 &   5.55E+02 &   3.43E+01 &   4.41E-02 &   1.01E-02 &     0.04 &     0.84 \\ 
 -2.40 &    4571 &    217 &   2.32E-11 &   1.35E-12 &   7.10E+02 &   4.11E+01 &   5.59E-02 &   1.30E-02 &     0.05 &     0.84 \\ 
 -2.20 &    4612 &    221 &   2.95E-11 &   1.72E-12 &   9.08E+02 &   4.98E+01 &   7.12E-02 &   1.67E-02 &     0.06 &     0.85 \\ 
 -2.00 &    4658 &    224 &   3.75E-11 &   2.16E-12 &   1.16E+03 &   6.05E+01 &   9.12E-02 &   2.14E-02 &     0.06 &     0.88 \\ 
 -1.80 &    4711 &    225 &   4.77E-11 &   2.68E-12 &   1.49E+03 &   7.46E+01 &   1.17E-01 &   2.75E-02 &     0.07 &     0.93 \\ 
 -1.60 &    4773 &    221 &   6.05E-11 &   3.27E-12 &   1.92E+03 &   9.27E+01 &   1.53E-01 &   3.50E-02 &     0.08 &     1.00 \\ 
 -1.40 &    4849 &    213 &   7.65E-11 &   3.94E-12 &   2.46E+03 &   1.17E+02 &   2.00E-01 &   4.41E-02 &     0.09 &     1.10 \\ 
 -1.20 &    4944 &    196 &   9.61E-11 &   4.83E-12 &   3.14E+03 &   1.52E+02 &   2.67E-01 &   5.44E-02 &     0.10 &     1.22 \\ 
 -1.00 &    5066 &    170 &   1.20E-10 &   6.20E-12 &   4.01E+03 &   2.05E+02 &   3.61E-01 &   6.57E-02 &     0.10 &     1.38 \\ 
 -0.80 &    5221 &    141 &   1.47E-10 &   8.30E-12 &   5.08E+03 &   2.83E+02 &   5.02E-01 &   8.04E-02 &     0.10 &     1.56 \\ 
 -0.60 &    5420 &    126 &   1.79E-10 &   1.12E-11 &   6.40E+03 &   3.89E+02 &   7.44E-01 &   1.24E-01 &     0.09 &     1.76 \\ 
 -0.40 &    5676 &    154 &   2.13E-10 &   1.57E-11 &   7.97E+03 &   5.27E+02 &   1.27E+00 &   3.48E-01 &     0.08 &     1.99 \\ 
 -0.20 &    6000 &    227 &   2.46E-10 &   2.40E-11 &   9.72E+03 &   7.63E+02 &   2.64E+00 &   1.20E+00 &     0.06 &     2.22 \\ 
  0.00 &    6412 &    332 &   2.70E-10 &   3.65E-11 &   1.14E+04 &   1.15E+03 &   6.71E+00 &   4.08E+00 &    -0.01 &     2.47 \\ 
  0.20 &    6919 &    481 &   2.82E-10 &   5.17E-11 &   1.28E+04 &   1.66E+03 &   1.98E+01 &   1.47E+01 &    -0.13 &     2.71 \\ 
  0.40 &    7500 &    645 &   2.85E-10 &   6.63E-11 &   1.39E+04 &   2.21E+03 &   5.69E+01 &   4.38E+01 &    -0.27 &     2.91 \\ 
  0.60 &    8084 &    759 &   2.83E-10 &   7.78E-11 &   1.49E+04 &   2.74E+03 &   1.33E+02 &   9.44E+01 &    -0.39 &     3.05 \\ 
  0.80 &    8619 &    813 &   2.81E-10 &   8.64E-11 &   1.58E+04 &   3.26E+03 &   2.52E+02 &   1.60E+02 &    -0.49 &     3.14 \\ 
  1.00 &    9086 &    817 &   2.79E-10 &   9.30E-11 &   1.67E+04 &   3.77E+03 &   4.08E+02 &   2.30E+02 &    -0.56 &     3.19 \\ 
  1.20 &    9478 &    782 &   2.79E-10 &   9.82E-11 &   1.76E+04 &   4.30E+03 &   5.90E+02 &   2.92E+02 &    -0.61 &     3.20 \\ 
  1.40 &    9797 &    719 &   2.82E-10 &   1.02E-10 &   1.86E+04 &   4.83E+03 &   7.82E+02 &   3.36E+02 &    -0.64 &     3.18 \\ 
  1.60 &   10060 &    644 &   2.87E-10 &   1.06E-10 &   1.97E+04 &   5.37E+03 &   9.84E+02 &   3.64E+02 &    -0.66 &     3.13 \\ 
  1.80 &   10290 &    573 &   2.96E-10 &   1.08E-10 &   2.09E+04 &   5.89E+03 &   1.21E+03 &   3.95E+02 &    -0.67 &     3.06 \\ 
  2.00 &   10493 &    509 &   3.09E-10 &   1.10E-10 &   2.25E+04 &   6.38E+03 &   1.46E+03 &   4.36E+02 &    -0.67 &     2.96 \\ 
\hline
\end{tabular}
\end{table*}

In short, the equations of mass, momentum and energy conservation
together with the simultaneous solution of the 3D radiative transfer
equation along nine inclined directions 
have been solved on a Eulerian mesh with 200\,x\,200\,x\,82 gridpoints:
for the spectral line formation calculation this was subsequently 
interpolated to a 50\,x\,50\,x\,82 grid. 
The physical dimension (6\,Mm horizontally and 3.8\,Mm vertically with about 1\,Mm
located above $\tau_{\rm 500} = 1$) 
of the numerical grid were sufficiently large
to cover $\ga 10$ granules simultaneously and extends to nearly adiabatic
conditions in the bottom. 
Periodic horizontal and open vertical boundary conditions have been employed.
The simulations make use of the state-of-the-art MHD equation-of-state
(Mihalas et al. 1988) and comprehensive opacities, including line opacities
(Gustafsson et al. 1975 with subsequent updates; Kurucz 1993).
The effects of line-blanketing is accounted for through opacity binning
(Nordlund 1982) which is a form of multi-group opacities. 
Some relevant properties of
the temporally and spatially (over surfaces of same optical depths)
averaged 3D solar simulation is listed in Table \ref{t:model}, limited
to the optical depth range relevant for spectrum formation. 
The mean temperature structure $<T(\tau_{500})>_\tau$ is shown in Fig. \ref{f:ttau}.
We emphasize, however, that the atmospheric inhomogeneities are also
important for the spectrum formation, which has been fully taken into
account in the 3D line formation.
The reader is referred to Stein \& Nordlund (1998) and Paper I for further
details of the construction of the 3D solar model atmosphere.

For comparison purposes, we have also performed identical calculations
employing two well-used 1D hydrostatic models of the solar atmosphere: the
semi-empirical Holweger-M\"uller (1974) model and a theoretical,
LTE, line-blanketed {\sc marcs} model (Asplund et al. 1997).
Their temperature structures $T(\tau_{500})$ are 
compared with the mean 3D stratification in Fig. \ref{f:ttau}.

\subsection{3D LTE spectral line formation}

Equipped with the 3D hydrodynamical solar model atmosphere, 3D spectral
line formation calculations have been performed for the [O\,{\sc i}], O\,{\sc i}
and OH lines. The radiative transfer equation has been solved under the assumption
of LTE for the ionization and excitation with the source function being
equal to the local Planck function ($S_\nu = B_\nu$).
Comprehensive background continuous opacities 
(Gustafsson et al. 1975; Asplund et al. 1997)
and an equation-of-state which accounts for excitation, ionization
and molecule formation of the most important elements (Mihalas et al. 1988)
have been used, with instantaneous chemical equilibrium
being assumed for the molecule formation.
For the [O\,{\sc i}] and O\,{\sc i} lines flux profiles have been
computed from a solution of the radiative transfer equation along
17 different inclined directions
($N_\mu = N_\varphi = 4$ plus the vertical) 
making use of the periodic horizontal boundaries. For the OH lines
disk-center intensity profiles have been obtained. The flux profiles
have been disk-integrated assuming a solar rotation velocity of
1.8\,km\,s$^{-1}$.
All 3D LTE spectral line formation computations have been carried out for 100 snapshots
covering 50\,min of solar time after which temporal and spatial averaging
have been performed. For each line, a local continuum flux/intensity has been
computed with the same procedure as for the line profiles, enabling proper
normalized profiles. Each line has been calculated with three different
abundances differing by 0.2\,dex from which the profile with the correct
equivalent width (OH lines) or line shape ([O\,{\sc i}] and O\,{\sc i} lines)
has been interpolated. Extensive testing has shown that the procedure
yields correct abundances to within 0.01\,dex.
No micro- and macroturbulence enter the 3D calculations as the self-consistently
computed convective velocities and temperature inhomogeneities
induce Doppler shifts which give rise to excellent agreement between
observed and predicted line shapes without such additional line broadening
which is always needed in standard 1D analyses
(Asplund et al. 2000b; Asplund 2000; Allende Prieto et al. 2002a).

The corresponding 1D calculations based on the {\sc marcs} and Holweger-M\"uller
model atmospheres have been performed using exactly the same codes and
procedures as for the 3D case, with the exception that a microturbulence of
$\xi_{\rm turb} = 1.0$\,km\,s$^{-1}$ has been adopted.
To investigate the sensitivities of the lines to this parameter,
all 1D profiles have also been computed using $\xi_{\rm turb} = 1.5$\,km\,s$^{-1}$.

\subsection{3D non-LTE spectral line formation for O\,{\sc i}
\label{s:nlte}}

It is well-known that the high-excitation permitted O\,{\sc i}
lines suffer from significant departures from LTE (Sedlmayr 1974; Kiselman 1991;
Kiselman \& Nordlund 1995; Kiselman 2001). It is therefore paramount
to investigate these non-LTE effects also in 3D and include non-LTE abundance
corrections in the final analysis of the O\,{\sc i} lines.
The 3D non-LTE line formation calculations have been performed with
{\sc multi3d} (Botnen 1997; Botnen \& Carlsson 1999; Asplund et al. 2003a),
which iteratively solves the rate equations simultaneously with
the radiative transfer equation assuming statistical equilibrium
(${\rm d}N_{\rm i}(x,y,z)/{\rm d}t = 0$).
Due to the very computationally demanding nature of 3D non-LTE line formation
calculations, compromises must be made in terms of the number of snapshots
for which the non-LTE solution is obtained. From the 50\,min time sequence for
which the 3D LTE line formation calculations were performed, two typical snapshots were
chosen which were sufficiently separated in time to be considered independent.
To improve the numerical accuracy in the non-LTE case, the 50\,x\,50\,x\,82 grid
used for the 3D LTE spectrum synthesis
were interpolated to a finer 50\,x\,50\,x\,100 grid extending down to
at least log\,$\tau_{500} = 2$ for all vertical columns.
Test calculations were also performed on a smaller 25\,x\,25\,x\,100 grid, yielding
very similar results.
Tests with different vertical depth-scales yielded insignificant differences
in the resulting non-LTE abundance corrections. 

\begin{figure}[t]
\resizebox{\hsize}{!}{\includegraphics{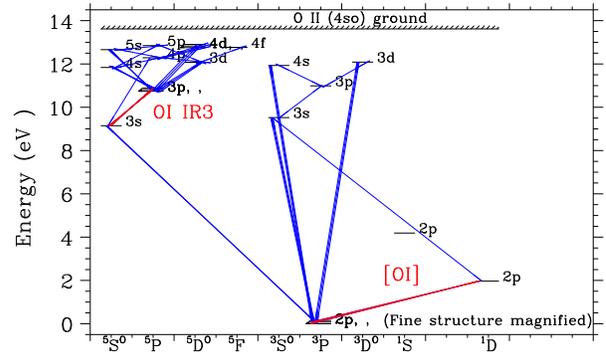}}
\caption{Grotrian term-diagram of the employed 23-level O model atom
with 22 bound states of O\,{\sc i} and the ground state of O\,{\sc ii}.
The forbidden [O\,{\sc i}] 630.0 and 636.3\,nm lines originate from
the 2p$^4$ $^3$P ground state, while the O\,{\sc i} 777\,nm triplet is between
the high-excitation 3s $^5$S$^o$ and 3p $^5$P levels. The 43 bound-bound
transitions are drawn. All O\,{\sc i} levels are connected by bound-free
transitions with the ground state of O\,{\sc ii} but are not shown for clarity
}
         \label{f:OI_atom}
\end{figure}

The radiative transfer equation was solved for 24 outgoing inclined rays
using a short characteristic technique and making use of the
periodic horizontal boundary conditions.
The line broadening caused by Doppler shifts introduced by the convective
motions in the hydrodynamical model atmospheres was taken into account.
No microturbulence or macroturbulence thus enter the 3D calculations.
The background continuous opacities and equation-of-state were computed with the
Uppsala opacity package (Gustafsson et al. 1975 and subsequent updates).
For completeness we note that this equation-of-state is not the
same as used in the construction of the 3D model atmosphere
but very similar in the relevant temperature regime. 
The effect of this difference will however be minute since 
both the LTE and non-LTE line profiles for the two selected 
snapshots used for the derivation of the non-LTE abundance corrections
are both based on the same input physics in all aspects,
as described in detail below. This non-LTE abundance correction is
then applied to the LTE estimates based on the whole 100 snapshot long
temporal sequence, which indeed have the same equation-of-state as
the 3D model atmosphere.
Finally, we note that all four codes employed in this study
(3D hydrodynamics, 3D LTE line formation, {\sc multi}
for 1D non-LTE line formation
and {\sc multi3d} for 3D non-LTE line formation) 
rely on the Uppsala opacity package for
continuous opacities.

Line-blanketing using data from Kurucz (1993)
were taken into account for the photo-ionization rates
but was found to be insignificant for the present case.
The main effect of the line blanketing on the statistical equilibrium
of metals that do not contribute significantly to the blanketing
is generally through the reduction of the available photoionizing 
UV photons, but with such a large ionization energy, oxygen is insensitive 
to changes in the UV solar flux 
(see, e.g. Allende Prieto et al. 2003).
Flux profiles were constructed from the angle-dependent intensity profiles.
The corresponding LTE-profiles for the selected snapshots were also produced with
{\sc multi3d} using the same model atom but with all collisional probabilities
multiplied by very large factors to ensure full thermalization of all rates.
The non-LTE abundance corrections were estimated by computing both the non-LTE and LTE
cases for three abundances differing by 0.2\,dex and interpolating the non-LTE 
line strengths
reproducing the LTE equivalent widths from the profile fitting. 
Both 3D snapshots gave the same non-LTE
correction to within 0.01\,dex. Similarly robust results were obtained in the case
of Li (Asplund et al. 2003a). Hence, while the line strengths vary somewhat
from snapshot to snapshot following differences in the granulation pattern, the non-LTE
abundance {\em corrections} stay essentially the same.

The O model atom consists of 22 O\,{\sc i} levels and the ground state
of O\,{\sc ii}. The term diagram of the adopted model atom is presented in
Fig. \ref{f:OI_atom}. In total 43 bound-bound and 22 bound-free radiative
transitions were included together with electron collisional excitation and
ionization.
Charge transfer reactions from the $^3$P ground states have been taken into 
account; large charge transfer cross-sections for excited states as
for Li\,{\sc i} (Barklem et al. 2003) are not expected for O.
We note that removing all charge transfer reactions only 
implies $<0.005$\,dex changes to the computed non-LTE abundance corrections
for the Sun, reflecting the very minor influence of all ionization rates.
Inelastic collisions with H for excitation and ionization were not accounted for,
as the available laboratory measurements and detailed calculations
strongly suggests that the classical Drawin (1968) formula over-estimates
the collisional rates by at least three orders of magnitude
in the cases of Li and Na
(Fleck et al. 1991; Belyaev et al. 1999; Barklem et al. 2003;
see also discussion in Kiselman 2001).
Unfortunately the specific case of O has not yet been investigated
but there are no compelling reasons on atomic physics grounds 
to expect a qualitatively different behaviour for O compared with Li and Na
in this respect. Furthermore,  center-to-limb variation
observations of the O\,{\sc i} triplet suggests that the Drawin recipe
over-estimates the collisional efficiency
(Allende Prieto et al., in preparation).
The adopted model atom is similar to the atom employed in the study
of Kiselman (1993).
We have verified that even an equivalent two-level atom yields very similar
non-LTE abundance corrections as the full O atom.
With the expected main non-LTE effect being photon losses in the line itself
for the IR triplet our adopted atom is extensive enough to produce reliable results
(Kiselman 1993).
The 3D non-LTE abundance corrections for the O\,{\sc i} lines employed in the
present study are listed in Table \ref{t:nlte}.

\begin{figure}[t]
\resizebox{\hsize}{!}{\includegraphics{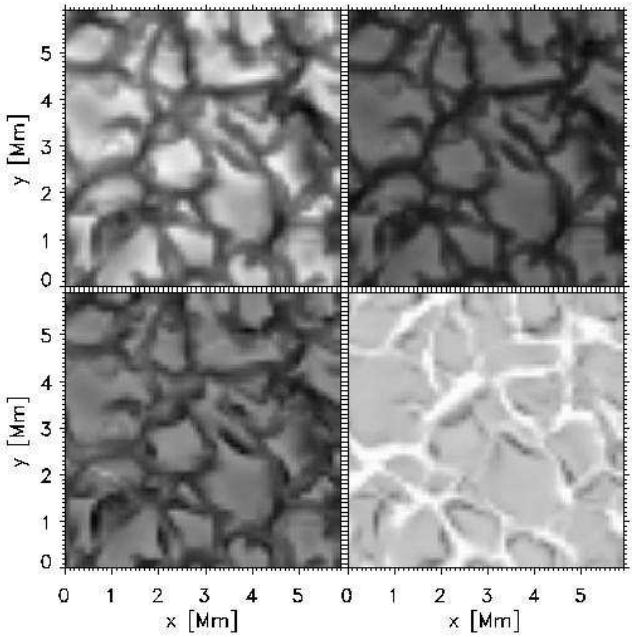}}
\caption{The granulation pattern in one of the solar snapshots
seen in disk-center continuum intensity at 777\,nm ({\it upper left panel}) and
equivalent width of the O\,{\sc i} 777.41\,nm line in
LTE ({\it upper right panel}) and non-LTE ({\it lower left panel});
the equivalent width images have the same relative intensity scale to
emphasize the overall difference in line strengths.
Also shown is the ratio of the non-LTE and LTE equivalent widths
({\it lower right panel}). 
}
         \label{f:OI_xy}
\end{figure}

\begin{figure}[t]
\resizebox{\hsize}{!}{\includegraphics{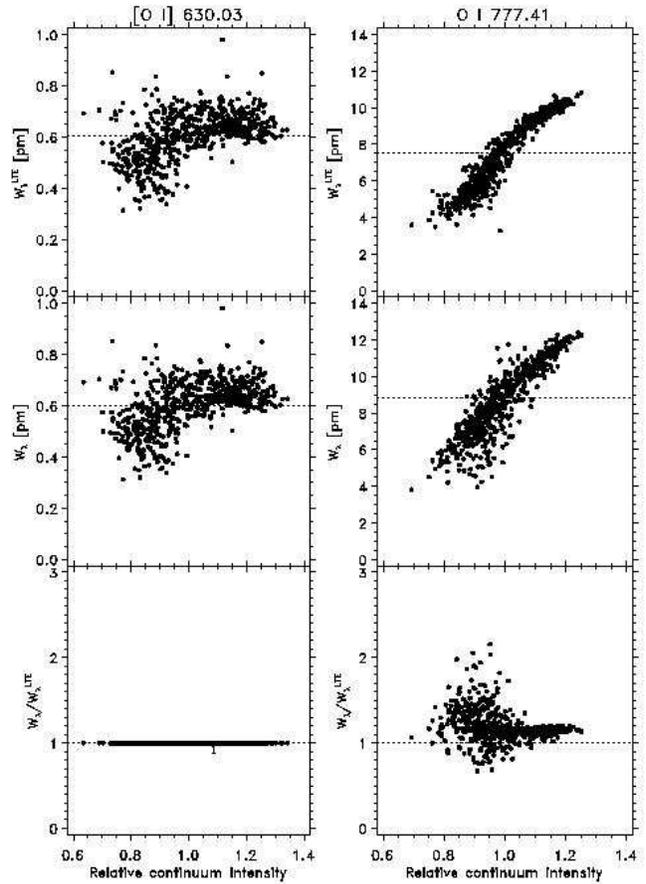}}
\caption{Predicted intensity ($\mu = 1.0$) equivalent widths of
the [O\,{\sc i}] 630.03\,nm ({\it left panel}) and O\,{\sc i} 777.41\,nm
({\it right panel}) lines across the granulation pattern in
LTE ({\it upper panel}) and non-LTE ({\it middle panel}). Both cases
are computed for log\,$\epsilon_{\rm O} = 8.90$ for the two lines.
The mean intensity equivalent widths are denoted by horizontal lines.
Also shown are the ratios of the non-LTE and LTE equivalent widths
({\it lower panel}), with the horizontal lines representing
the average ratio for the two lines.
As expected, the [O\,{\sc i}] line is in perfect LTE
while the O\,{\sc i} line shows significant departures from LTE.
Note that for clarity only a selection of the vertical atmospheric 
columns are shown.
}
         \label{f:OIvsI}
\end{figure}

Due to their high excitation potential ($\chi_{\rm exc} > 9$\,eV),
the O\,{\sc i} lines are formed in deep atmospheric layers with a significant
temperature contrast between the warm upflowing gas (granules) and the cool downflows
(intergranular lanes), as exemplified in Fig. \ref{f:OI_xy}.
It is clear that the main effect of the non-LTE line formation is a general
strengthening of the O\,{\sc i} IR triplet over the whole granulation pattern.
In this sense, the non-LTE line formation is largely a 1D problem: quite similar
results are obtained for O\,{\sc i} if the non-LTE calculations are restricted
to 1.5D (i.e. ignoring all non-vertical radiative transfer) rather than 3D.
The depression of the line source function thus operates qualitatively
similarly in both up- and downflows, although the details differ somewhat.
As seen in Fig. \ref{f:OI_xy}, the largest 3D non-LTE effects are in fact
seen in the intergranular lanes, which however, also house the smallest
non-LTE effects in regions immediately adjacent to edge-brightened granules.
Fig. \ref{f:OIvsI} shows the line strength of the O\,{\sc i} 777.41\,nm line
as a function of the local continuum intensity. Again, the non-LTE and LTE
behaviour are qualitatively similar for the triplet with the main difference
being a shift in overall equivalent widths for the same abundance.
The lack of a clear difference between the two cases
in this respect is in sharp contrast to the case of Li
(Kiselman 1997: Uitenbroek 1998; Asplund et al. 2003a), where the shapes of the
LTE and non-LTE loci differ markedly.
It will therefore be difficult to confirm the 3D non-LTE
line formation modelling through observations of spatially resolved
O\,{\sc i} lines, as is possible for elements like Li and Fe
(Kiselman \& Asplund 2001).
As expected, the [O\,{\sc i}] lines are formed perfectly in LTE, as seen
in Fig. \ref{f:OIvsI}.

\begin{table}[t!]
\caption{The computed non-LTE abundance corrections for the
three different model atmospheres. The departures from LTE have
been estimated using {\sc multi3d} (Botnen 1997 Asplund et al. 2003a)
for the 3D case and with {\sc multi} (Carlsson 1986) for the 1D model atmospheres,
in all cases using the same 23-level model atom.
\label{t:nlte}
}
\begin{tabular}{lcccccc}
 \hline
line   &   	&	$\chi_{\rm exc}$ & log\,$gf$ &
\multicolumn{3}{c}{$\Delta({\rm log} \epsilon_{\rm O})$} \\
\cline{5-7}
$$[nm] & 	&	[eV]             &           &   3D & HM & {\sc marcs} \\
 \hline
[O\,{\sc i}]: \\
630.03 &  $^3$P -- $^1$D & 0.00 & $-9.72$  & $+0.00$ & $+0.00$ & $+0.00$ \\
636.67 &  $^3$P -- $^1$D & 0.02 & $-10.19$ & $+0.00$ & $+0.00$ & $+0.00$ \\
O\,{\sc i}: \\
615.81 &  $^5$P -- $^5$D$^{\rm o}$ & 10.74 & $-0.30$ & $-0.03$ & $-0.05$ & $-0.03$ \\
777.19 &  $^5$S$^{\rm o}$ -- $^5$P &  9.15 & $+0.37$ & $-0.27$ & $-0.29$ & $-0.24$ \\
777.41 &  $^5$S$^{\rm o}$ -- $^5$P &  9.15 & $+0.22$ & $-0.24$ & $-0.27$ & $-0.23$ \\
777.53 &  $^5$S$^{\rm o}$ -- $^5$P &  9.15 & $+0.00$ & $-0.20$ & $-0.24$ & $-0.20$ \\
844.67 &  $^3$S$^{\rm o}$ -- $^3$P &  9.52 & $+0.01$ & $-0.20$ & $-0.25$ & $-0.21$ \\
926.60 &  $^5$P -- $^5$D$^{\rm o}$ & 10.74 & $+0.82$ & $-0.08$ & $-0.11$ & $-0.08$ \\
\hline
\end{tabular}
\end{table}

The 1D non-LTE line formation with the {\sc marcs} and the Holweger-M\"uller
model atmospheres has been studied with the code {\sc multi} (Carlsson 1986)
using the same model atom as in the 3D case. The structures of the
{\sc multi3d} and {\sc multi} codes are very similar with the exceptions of
the geometry and the use of long characteristics for the solution of
the radiative transfer with {\sc multi}. In particular, the subroutines for the
background opacities and equation-of-state calculations are identical in the two cases.
The non-LTE abundance corrections
for the two 1D model atmospheres are presented in Table \ref{t:nlte}.
It is worth noting the similarity between the non-LTE corrections in 1D and 3D
in the case of O\,{\sc i} lines in the Sun, in particular for the theoretical
{\sc marcs} model. 
Although detailed results require 3D calculations, non-LTE computations
in 1D model atmospheres are of great value to explore whether a given spectral
line is prone to departures from LTE, at least at solar metallicities.
At low metallicities, the temperature structure is sufficiently different
between 1D and 3D model atmospheres that one can expect significantly 
larger non-LTE effects in 3D for many elements (Asplund et al. 1999a, 2003a).

\section{The solar photospheric O abundance
\label{s:results}}

\subsection{Forbidden [O\,{\sc i}] lines}

The ability of the 3D hydrodynamical solar
model atmosphere to very accurately predict detailed spectral line shapes
and asymmetries without invoking additional ad-hoc broadening like
micro- and macroturbulence represents a major advantage in abundance analyses,
as obvious from the analysis of the forbidden [O\,{\sc i}] 630.0\,nm line.
It has long been suspected that the line is indeed blended by a
weak Ni\,{\sc i} line (Lambert 1968). Due to the until recently
very uncertain transition probability of the Ni line, it has been difficult to ascertain
its overall contribution to the 630.0\,nm feature and hence the solar
O abundance. Lambert (1978) estimated the Ni contribution to be
0.02\,pm=0.2\,m\AA , while the
use of the theoretical $gf$-value (log\,$gf = -1.74$) of Kurucz (1993) for the
Ni line would imply a very
small left-over contribution for the [O\,{\sc i}] line using the solar Ni
abundance derived from other isolated lines.
The 3D model atmosphere can be used to estimate the necessary Ni line 
contribution in order to reproduce the shape of the observed feature.
Without taking into account the Ni line,
the predicted [O\,{\sc i}] line has completely the wrong central wavelength and
a line shape different from observed (see Fig. 1 in
Allende Prieto et al. 2001), a discrepancy which is far worse than
seen in similar 3D calculations for any other unblended line.

From a $\chi^2$-analysis of the flux profile of the [O\,{\sc i}] 630.0\,nm line,
Allende Prieto et al. (2001) concluded that
the solar O abundance is log\,$\epsilon_{\rm O} = 8.69 \pm 0.05$
based on the same 3D model atmosphere as employed here.
Since then, Johansson et al. (2003) have measured a new $gf$-value for
the blending Ni\,{\sc i} line and details of its isotopic
splitting. 
Johansson et al. have modelled the shape of the Ni\,{\sc i} line at the lab
with two major components, due to the isotopes $^{60}$Ni and $^{58}$Ni, which
contribute 94.3 \% of the {\it terrestrial} abundance (Rosman \& Taylor 1998).
Although the exact $gf$-value is inconsequential for our
purposes (only the product $\epsilon_{\rm Ni} \cdot gf$
enters the analysis\footnote{In principle, one should re-determine the solar Ni
abundance using other Ni lines with our new 3D model atmosphere in order to
evaluate whether the new $gf$-value for the Ni\,{\sc i} 630.0\,nm line is indeed consistent
with the solar Ni abundance. We postpone such a study to a later time as it is not
crucial for our conclusions here.}),
the additional line broadening from isotopic splitting
changes the line profile of the Ni line somewhat and
therefore also slightly the $\chi^2$-analysis. 
The center of gravity of the line is also mildly changed from the value
used by Allende Prieto et al. (630.0339 nm; Litz\'en et al. 1993)
to 630.0341 nm.
Taking this into account, however, only results in a 0.005\,dex higher abundance
than without the isotopic splitting. 
We therefore still arrive at
log\,$\epsilon_{\rm O} = 8.69 \pm 0.05$ (Table \ref{t:OI}).
The same abundance is obtained when using disk-center intensity profiles
instead of flux profiles.

Adopting the theoretical line strength of only the [O\,{\sc i}] 630.0\,nm line
as judged by the 3D analysis (0.36\,pm=3.6\,m\AA\ for disk-center intensity
and 0.43\,pm for flux) for
the 1D calculations using the Holweger-M\"uller and the {\sc marcs} model
atmospheres, we find log\,$\epsilon_{\rm O} = 8.77 \pm 0.05$ and
log\,$\epsilon_{\rm O} = 8.73 \pm 0.05$, respectively. It should be noted,
however, that it is not possible to accurately evaluate the Ni line
contribution to the feature based only on a 1D analysis (Lambert 1968;
Reetz 1998) due to the absence of intrinsic line asymmetries
and the poor overall agreement between predicted
and observed line profiles with 1D hydrostatic models.
For a long time, it was therefore simply assumed that the Ni line was
unimportant, leading to a much larger derived solar O abundance
(e.g. Lambert 1978; Sauval et al. 1984):
log\,$\epsilon_{\rm O} \simeq 8.90$ for the Holweger-M\"uller model
using the new $gf$-value of the [O\,{\sc i}] 630.0\,nm line adopted here.
This clearly illustrates the importance of having a realistic model
atmosphere which adequately describes the intrinsic line shapes and
asymmetries due to the convective motion and temperature inhomogeneities.

As mentioned above, the analysis of the [O\,{\sc i}] 636.3\,nm line is hampered by
blending CN lines. In addition, the line is located in the midst of a Ca\,{\sc i}
auto-ionization line. The main blending CN line is the Q$_2$ (25.5) 10-5
line with a center wavelength of 636.3776\,nm, which is very close to
the wavelength of the [O\,{\sc i}] line: 636.3792\,nm. In addition, the
P$_1$ (53.5) 8-3 CN line at 636.3846\,nm is blending in the red wing.
The contributions of these CN lines have been determined from spectrum
synthesis using the parameters for the two bands determined from other
CN lines as well as for theoretical values. Together the two CN lines
are estimated to have an equivalent width of $0.050 \pm 0.005$\,pm (flux),
where the uncertainty reflects the different possible choices of
transition probabilities of the CN lines. This leaves
$0.14$\,pm (1.4\,m\AA ) to be attributed to the [O\,{\sc i}] line.
In terms of abundance this corresponds to log\,$\epsilon_{\rm O} = 8.67$
when analysed with the 3D hydrodynamical solar model atmosphere
(Table \ref{t:OI}).
For the Holweger-M\"uller and the {\sc marcs} model
atmospheres, we find log\,$\epsilon_{\rm O} = 8.75$ and
log\,$\epsilon_{\rm O} = 8.71$, respectively.

The [O\,{\sc i}] 557.73\,nm line is not considered here since it is badly
blended by a C$_2$ doublet whose exact contribution is very difficult
to ascertain (Lambert 1978).

\subsection{Permitted O\,{\sc i} lines}

Of the 20 odd observable lines attributed to O\,{\sc i} in the wavelength range
$500-1400$\,nm, unfortunately only a few are suitable for accurate
abundance analysis besides the two forbidden lines discussed above.
The O\,{\sc i} 615.68\,nm line is definitely blended as are five lines
at wavelengths greater than 900\,nm (926.27, 974.15,
1130.24, 1316.39 and 1316.49\,nm). The heavy blending preventing their use
has been verified by detailed 3D spectrum synthesis, as the shapes of clean lines
are very well predicted by the new generation of 3D hydrodynamical model
atmosphere (e.g. Asplund et al. 2000b; Asplund 2000; Allende Prieto et al. 2002a).
The two very faint ($W_\lambda < 0.1$\,pm) lines at 645.44
and 700.19\,nm are not considered here because of their weakness and their
uncertain theoretical $gf$-values.
The 844.63\,nm line is affected by uncertainties in the continuum
placement and blending while the 976.07\,nm lacks appropriate atomic data.
This leaves only the familiar O\,{\sc i} IR triplet at 777\,nm
(777.19, 777.41, 777.53\,nm) and the even more high-excitation
615.81\,nm line. In addition, we include the 844.67\,nm line but with
half weight compared with the other lines
as it is partly blended and located in a region with
a less well-determined continuum. The latter line also shows similarly large
non-LTE abundance corrections as the IR triplet, i.e. $\simeq -0.2$\,dex.
Finally, we include the 926.60\,nm line, also with half weight as the
line is blended in the red wing by a telluric H$_2$O line. As a consequence,
it is not possible to use the flux atlas to derive the abundance.
While the Brault \& Neckel (1987) disk-center intensity solar 
atlas\footnote{ftp.hs.uni-hamburg.de/pub/outgoing/FTS-atlas}
still shows a prominent telluric feature, the Li\`ege Jungfraujoch
(Delbouille et al. 1973) disk-center 
atlas\footnote{http://bass2000.obspm.fr/solar\_spect.php}
has a diminished problem with telluric absorption in this
wavelength region. For this line only we therefore fit the disk center
intensity profile instead of the flux profile for this particular line. 
Note that the line
consists in fact of three components at 926.5827, 926.5927 and 926.6007\,nm
with log\,$gf = -0.72$, $+0.13$ and $+0.71$, respectively, which
have been taken into account for the theoretical calculations.

The derived O\,{\sc i}-based abundances have been obtained from the
temporally and spatially averaged 3D LTE flux profiles and taking into
account the departures from LTE calculated in Sect. \ref{s:nlte}.
We note that almost identical abundances are derived when relying on
disk-center intensity profiles instead.
The 3D LTE profiles have been computed for 100 snapshots corresponding
to 50\,min solar time for in total 17 different rays.
Through a comparison with the observed profiles, the individual abundances best
reproducing the line shapes and strengths for the different O\,{\sc i} lines
were estimated. To these 3D LTE abundances, the 3D non-LTE abundance corrections
presented in Table \ref{t:nlte} are added.
The final 3D non-LTE abundances are found in Table \ref{t:OI}.
Due to the in general poorer agreement between predicted and observed profiles
with 1D model atmospheres, the same profile fitting technique is not as easily employed
in the 1D cases. Instead, the theoretical equivalent widths in 3D computed with the
thus obtained abundances have been used as ``observed" equivalent widths to
be reproduced with the two different 1D model atmospheres. This ensures
that the same line strengths are obtained in both 1D and 3D and isolates
the impact of the new generation of 3D solar model atmosphere compared with
existing 1D models. As in 3D, the 1D non-LTE abundance corrections
individually computed for the two 1D model atmospheres have been added to the
results presented in Table \ref{t:OI}.

\begin{table}[t!]
\caption{The derived solar oxygen abundance as indicated by
forbidden [O\,{\sc i}] and permitted O\,{\sc i} lines. The results
for the O\,{\sc i} lines include the non-LTE abundance corrections presented
in Table \ref{t:nlte} which have been computed specifically for the
three different model atmospheres
\label{t:OI}
}
\begin{tabular}{lccccccc}
 \hline
line   & $\chi_{\rm exc}$ & log\,$gf$ & 3D$^{\rm a}$ &
$W_\lambda^{\rm b}$ & HM$^{\rm c}$ & {\sc marcs}$^{\rm c}$ \\
$$[nm] & [eV]             &           &              &
[pm]                &              &                       \\
 \hline
[O\,{\sc i}]: \\
630.03 & 0.00 & $-9.72$  & 8.69 & 0.43 & 8.77 & 8.73 \\
636.37 & 0.02 & $-10.19$ & 8.67 & 0.14 & 8.75 & 8.71 \\
O\,{\sc i}: \\
615.81 & 10.74 & $-0.30$ & 8.62 & 0.41 & 8.77 & 8.77 \\
777.19 &  9.15 & $+0.37$ & 8.64 & 7.12 & 8.60 & 8.71 \\
777.41 &  9.15 & $+0.22$ & 8.65 & 6.18 & 8.60 & 8.71 \\
777.53 &  9.15 & $+0.00$ & 8.66 & 4.88 & 8.62 & 8.71 \\
844.67$^{\rm d}$ &  9.52 & $+0.01$ & 8.60 & 3.52 & 8.58 & 8.67 \\
926.60$^{\rm e}$ &  10.74 & $+0.82$ & 8.65 & 3.62 & 8.68 & 8.69 \\
\hline
\end{tabular}
\begin{list}{}{}
\item[$^{\rm a}$] The abundances derived using the 3D model atmosphere have
been obtained from profile fitting of the observed lines.
\item[$^{\rm b}$] The predicted flux equivalent widths (except for 926.60\,nm
which is for disk center intensity) with the 3D model atmosphere
using the best fit abundances shown in the fourth column.
\item[$^{\rm c}$] The derived abundances with the Holweger-M\"uller (1974)
and the {\sc marcs} (Asplund et al. 1997) 1D hydrostatic model atmospheres
in order to reproduce the equivalent widths presented in the fifth column.
Note that for the [O\,{\sc i}] 630.03\,nm line use of 1D models prevents
accurate estimation of the significant Ni\,{\sc i} line blending
(Allende Prieto et al. 2001). Without the 3D results at hand, a significantly
larger O abundance of log\,$\epsilon_{\rm O} \simeq 8.9$ would therefore
be obtained.
\item[$^{\rm d}$] Half weight compared with the other lines is assigned
to the 844.67\,nm line due to problems with blending and continuum placement.
\item[$^{\rm e}$] Half weight compared with the other lines is assigned
to the 926.60\,nm line due to problems with blending and continuum placement.
Note that the line in fact consists of three components, which has been
taken into account in the spectrum synthesis. In order to minimize the
effect of the telluric H$_2$O blend, the disk-center intensity profile
has been fitted, and hence the tabulated equivalent width is for the intensity
rather than for flux profile as for all other lines here.
\end{list}

\end{table}

\begin{figure}[t]
\resizebox{\hsize}{!}{\includegraphics{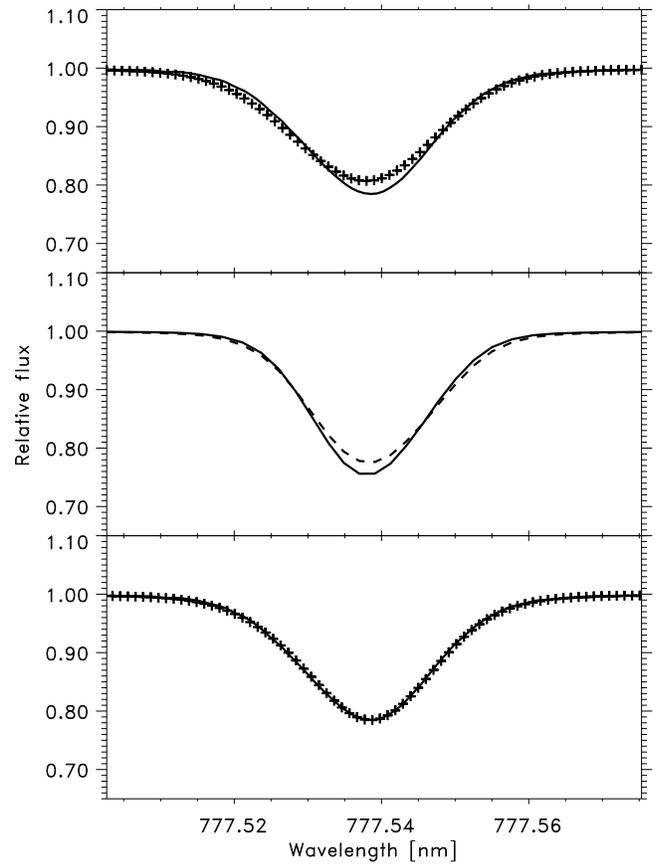}}
\caption{{\em Upper panel:} The temporally and spatially averaged LTE flux profile
from the 3D model atmosphere ($+$) together with the observed profile (solid line)
for the O\,{\sc i} 777.53\,nm line. The theoretical profile has been computed with
log\,$\epsilon_{\rm O} = 8.86$ and convolved
with a Gaussian corresponding to the instrumental resolution. Clearly the 3D LTE
profile lacks effects of departures from LTE in the line formation.
{\em Middle panel:} The 3D non-LTE (solid line) and the 3D LTE (dashed line)
line profiles
computed for one snapshot with {\sc multi3d} (Botnen 1997 Asplund et al. 2003a).
The 3D non-LTE profile has been computed with log\,$\epsilon_{\rm O} = 8.70$
while for the LTE case an abundance of log\,$\epsilon_{\rm O} = 8.90$ was adopted.
In terms of equivalent widths the two profiles have the same line strengths.
{\em Lower panel:} The temporally and spatially averaged 3D profile taking into
account departures from LTE ($+$) shown together with the observed profile (solid line).
The 3D non-LTE profile has been obtained by multiplying the 3D LTE profile in
the upper panel with the ratio of non-LTE and LTE profiles presented in the middle panel.
While not a perfect substitute for the temporally averaged 3D non-LTE profile
it nevertheless shows a very encouraging agreement with observations.
The observed line profiles have been corrected for the solar gravitational redshift
of $633$\,m\,s$^{-1}$.
}
         \label{f:OI_prof}
\end{figure}

It is noteworthy that the 3D non-LTE abundances in Table \ref{t:OI} show very
gratifying agreement in spite of the very significant differences in
non-LTE abundance corrections between the different lines. In particular,
the 615.81\,nm and 926.60\,nm lines 
show a small departure from LTE ($-0.03$\,dex and $-0.08$\,dex, respectively, in terms
of abundance) while the other four O\,{\sc i} lines have substantial non-LTE effects
($-0.20...-0.27$\,dex), yet the final non-LTE abundances agree to within 0.06\,dex.
While the non-LTE effect on the 615.81\,nm line is small, the 3D LTE effect
(3D LTE - 1D LTE) is
significantly larger than for the other O\,{\sc i} lines in view of its very
high excitation potential ($\chi_{\rm exc} = 10.74$\,eV) and large
atmospheric formation depth.
In sharp contrast, with the Holweger-M\"uller model atmosphere
the 615.81\,nm line implies a significantly larger abundance than the 777\,nm 
triplet by $\ga 0.15$\,dex. With the {\sc marcs} model atmosphere the agreement is
better but still significantly poorer than in 3D. 
It should be noted however that the here performed
1D analyses benefit greatly from the 3D analysis.
The equivalent widths estimated from the 3D line profile
fitting are much more accurate than those directly measured from 
the observed profiles using for example Gaussian or Voigt functions as
commonly done due to the asymmetries of the lines and the presence
of blends. 
We interpret the excellent
agreement in 3D as yet additional evidence for a very high degree of realism
in the 3D model atmosphere and 3D line formation modelling.

Not only are the derived abundances very consistent in 3D, the predicted
profiles also closely resemble the observed profiles,
as evident from Fig. \ref{f:OI_prof}.
While the 3D LTE profiles are broader and have a too pronounced blue
asymmetry compared with what the observations indicate, including the 3D non-LTE
effects results in almost perfect matches with the observed profiles
for the O\,{\sc i} IR triplet.
It should be noted that due to computational constraints the 3D non-LTE calculations
can only be performed for a small number of snapshots, preventing proper
temporally averaged 3D non-LTE profiles as can be straight-forwardly achieved
in 3D LTE. Instead the effects of departures from LTE on the 3D profiles have been
included by multiplying the temporally averaged 3D LTE profiles with
the ratios of the 3D non-LTE and 3D LTE profiles computed
with {\sc multi3d} for the two selected snapshots. This procedure is justified
since the differences in
3D non-LTE abundance corrections between different snapshots are expected to
be small (Asplund et al. 2003a), as also confirmed by specific calculations for O.
Thereby the final 3D non-LTE profiles incorporates the effects of the
Doppler shifts introduced by convective motions and solar oscillations, which
in 1D is assumed to be replaced with a depth-independent and isotropic
micro- and macroturbulence.
The agreement between predicted and observed profiles is in general significantly
poorer in 1D even when including an individually optimized macroturbulence parameter.
The excellent agreement between observed and predicted profiles for the
3D non-LTE calculations is a very strong support for our modelling.

\begin{figure}[t]
\resizebox{\hsize}{!}{\includegraphics{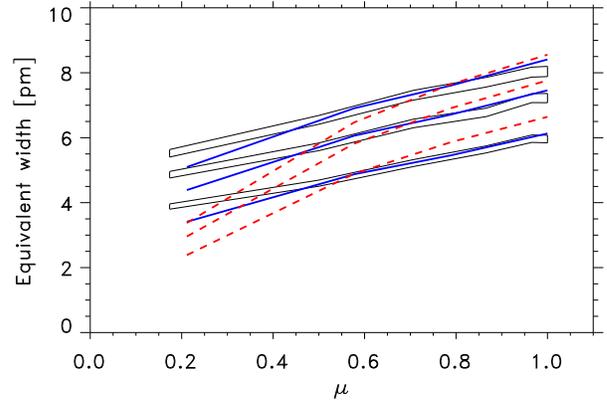}}
\caption{The observed center-to-limb variations of the equivalent widths
of the O\,{\sc i} 777\,nm triplet are shown as
bands with estimated internal errors. 
The solid lines denote the average
of two snapshots of the theoretical 3D non-LTE results for the three lines,
while the dashed lines represent the corresponding
3D LTE case. The here shown 3D LTE curves are very similar to the average of all 100
snapshots computed with the same 3D LTE line formation code used for
the calculations of the [O\,{\sc i}] and OH lines. As two snapshots
are insufficient to yield an accurate abundance estimate, the 3D non-LTE results
have been interpolated to log\,$\epsilon_{\rm O} \approx 8.7$ to
overlap with the observational estimates to high-light the excellent agreement
in the center-to-limb variation in 3D non-LTE. In all cases, the LTE curves correspond
to a 0.2\,dex higher abundance than the non-LTE cases. Clearly,
LTE fails to reproduce the observational evidence, while
the non-LTE calculations are in good agreement.
}
         \label{f:OI_limb}
\end{figure}

Further support for the 3D model atmosphere and the 3D non-LTE line
formation modelling is obtained from a comparison of the center-to-limb
variation of the O\,{\sc i} 777\,nm triplet.
As discussed in detail in Kiselman \& Nordlund (1995), the observed
dependence of line strength on disk location $\mu$  
is much more shallow
than predicted when assuming LTE regardless of whether a 3D or 1D model
atmosphere is employed.
In contrast, the non-LTE case reproduces the observed behaviour very nicely,
at least when not including the very large cross-sections for inelastic
collisions with H implied by the classical Drawin (1968) formula
(see discussion in Kiselman 2001).
Not surprisingly given the similarity in the employed 3D solar model atmospheres
and the robustness of the non-LTE effects as discussed in Sect. \ref{s:nlte},
we confirm the findings of Kiselman \& Nordlund (1995). In this respect,
our much more extended model atom compared with their two-level approach is
of little consequence, as the dominant non-LTE effect is a depression of
the line source function due to escape of 777\,nm line photons
(Kiselman 1993).
As shown in Fig. \ref{f:OI_limb}, the 3D non-LTE
results for the center-to-limb variations of the triplet
agree very well with the observational evidence,
in particular for $\mu \ge 0.5$.
We note that an even more impressive agreement for $\mu < 0.5$ would have been 
achieved had we relied on the only, to our knowledge at least, previously
published center-to-limb variation observations of the triplet by 
M\"uller et al. (1968) and Altrock (1968).
Our new solar observations are in excellent accordance with those of
M\"uller et al. (1968) and Altrock (1968), except for the data points at
$\mu =0.17$, the only disk position with $\mu < 0.5$ we acquired spectra for.

\subsection{OH vibration-rotation lines
\label{s:OHvr}}

The availability of high-resolution solar atlases in the IR with
exquisite signal-to-noise ratio such as the here employed ATMOS atlas
enables the use of the OH vibration-rotation around 3\,$\mu$m
and the OH pure rotation
lines beyond 9\,$\mu$m as oxygen abundance indicators
(Grevesse et al. 1984; Sauval et al. 1984).
The high-quality observed spectrum combined with the relatively small amount
of problematic blending in the relevant wavelength regions ensures that
highly precise equivalent widths (initially in milli-Kayser = milli-cm$^{-1}$
due to the nature of the solar atlas and subsequently converted to pm)
of the OH lines can be measured.
For the analysis of the OH $X^2\Pi$ vibration-rotation lines we utilise
70 lines in the $P_i$-branch of the (1,0) and (2,1) bands with $N'' < 20$;
the $Q$ and $R$ lines fall in wavelength regions completely dominated
by CO$_2$ and H$_2$O. The (1,0) and (2,1) lines are all weak and hence
insensitive to the non-thermal broadening (convective motions in 3D
and microturbulence in 1D). Due to the large number of lines,
our OH-based oxygen abundances are determined using
the measured equivalent widths with the assumption of LTE in the
line formation.
For vibration-rotation lines like these, LTE is likely an excellent
approximation (Hinkle \& Lambert 1975). The issue of non-equilibrium
molecule formation for OH has garnered relatively little attention
in the literature but the available evidence suggests that
it is not a serious worry for the OH lines in the Sun at least
(S\'anchez Almeida et al. 2001; Asensio Ramos et al. 2003).

\begin{figure}[t]
\resizebox{\hsize}{!}{\includegraphics{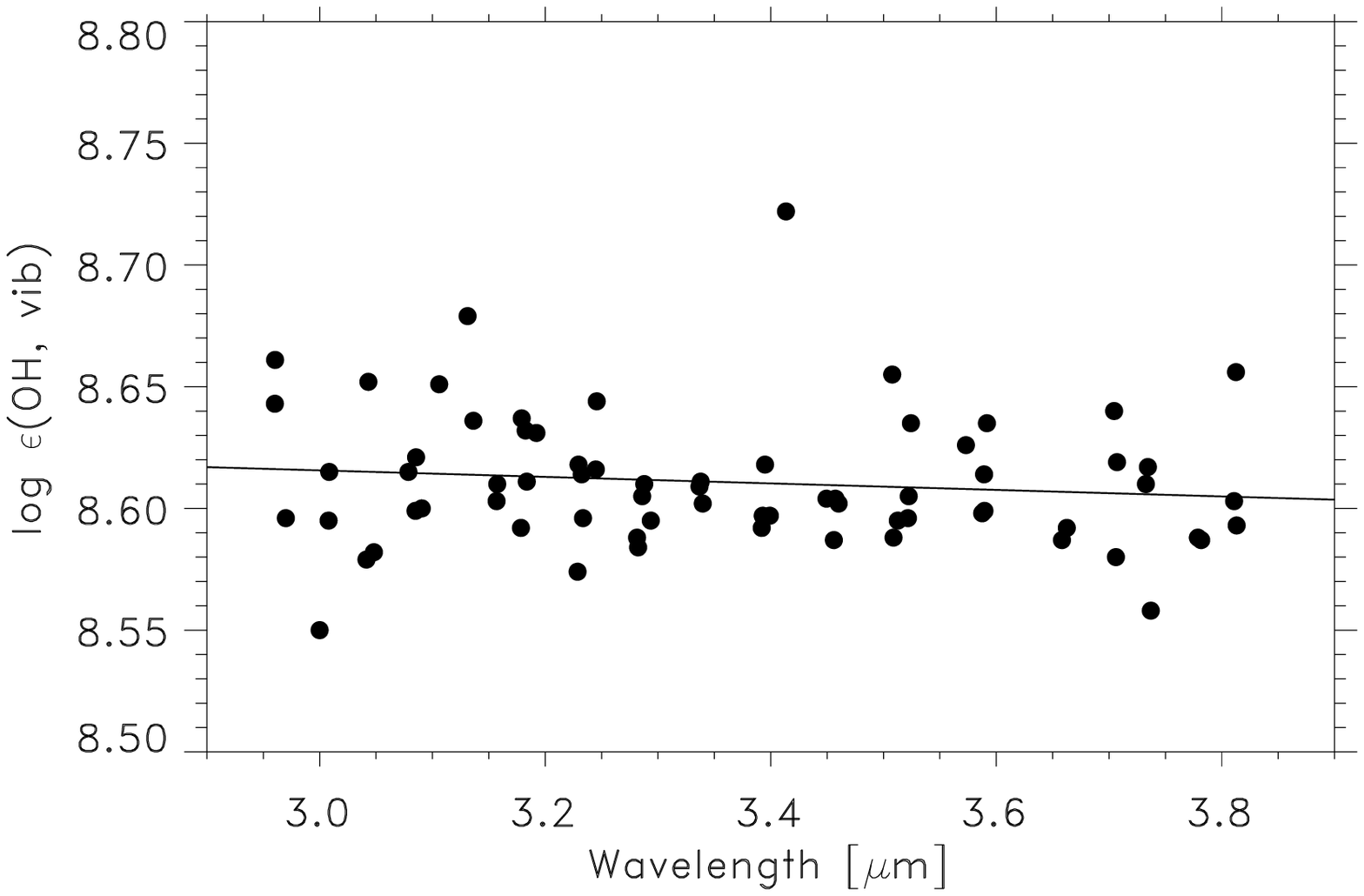}}
\resizebox{\hsize}{!}{\includegraphics{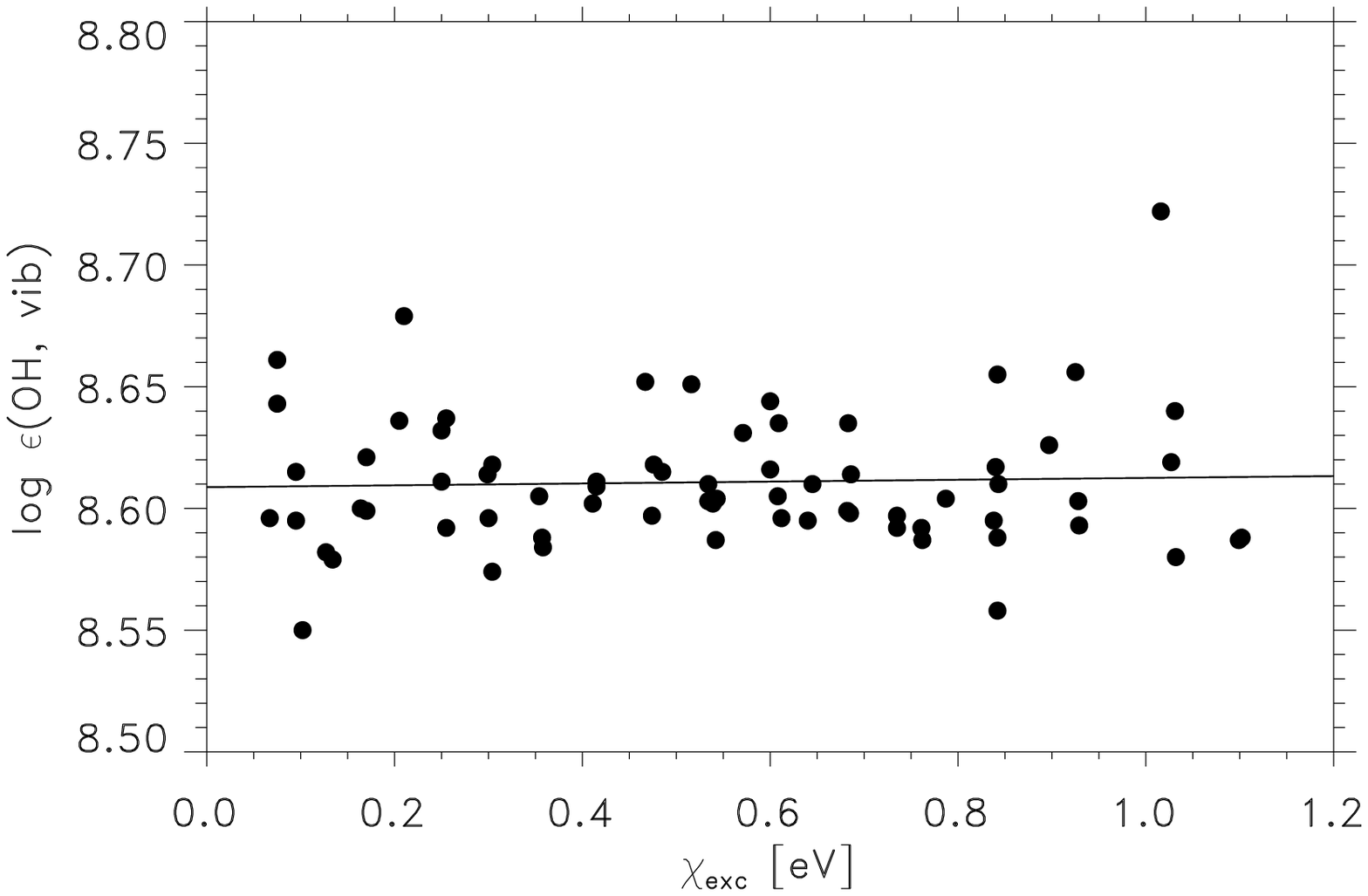}}
\resizebox{\hsize}{!}{\includegraphics{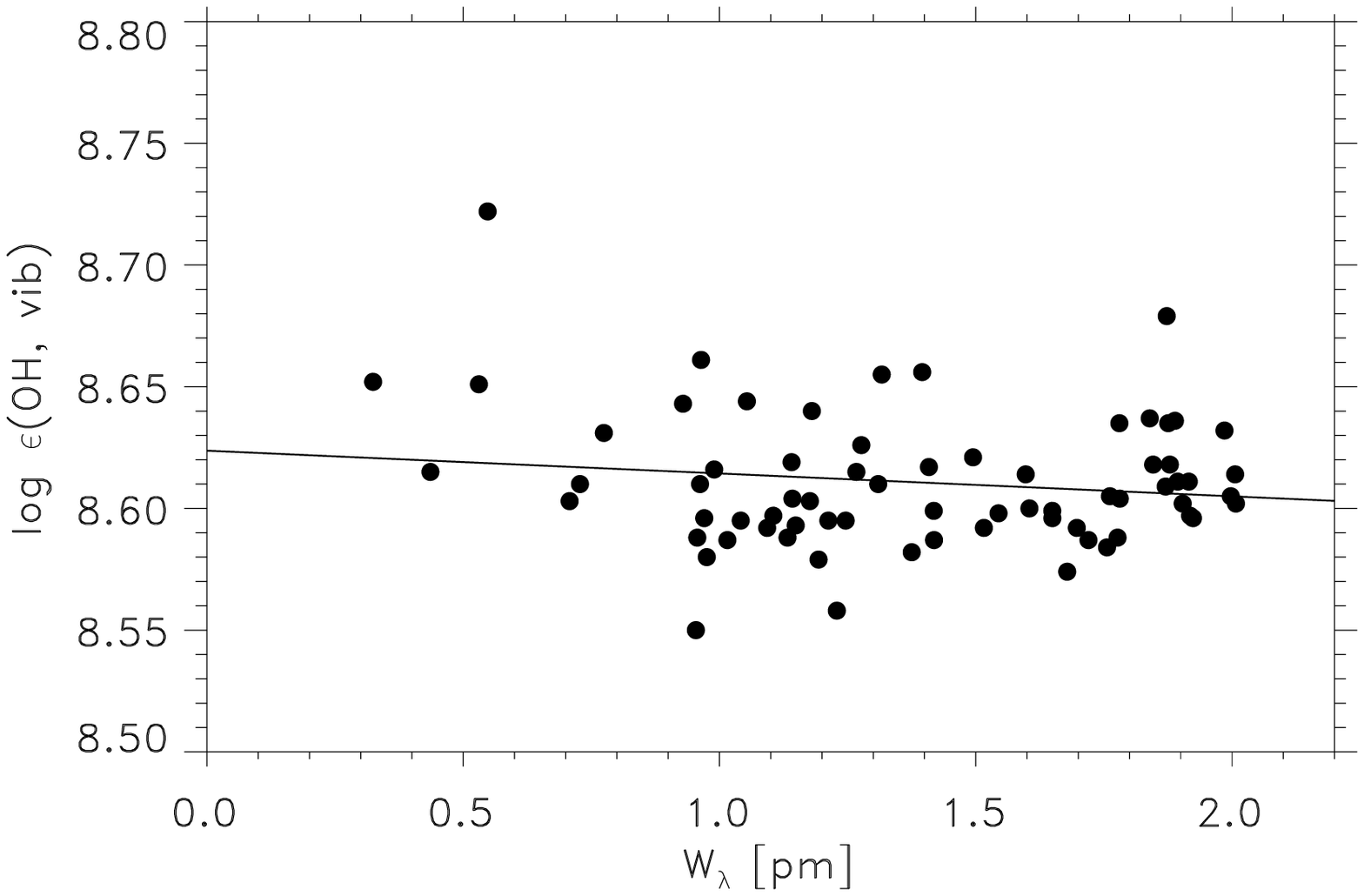}}
\caption{The derived solar oxygen abundance (filled circles)
from OH vibration-rotation
lines using the 3D hydrodynamical
time-dependent simulation of the solar atmosphere (Asplund et al. 2000b)
as a function of wavelength, lower level excitation
potential and line strength (in pm).
The solid lines denote least-square-fits giving equal weights to all lines. 
}
         \label{f:OHvr_3D}
\end{figure}

\begin{figure}[t]
\resizebox{\hsize}{!}{\includegraphics{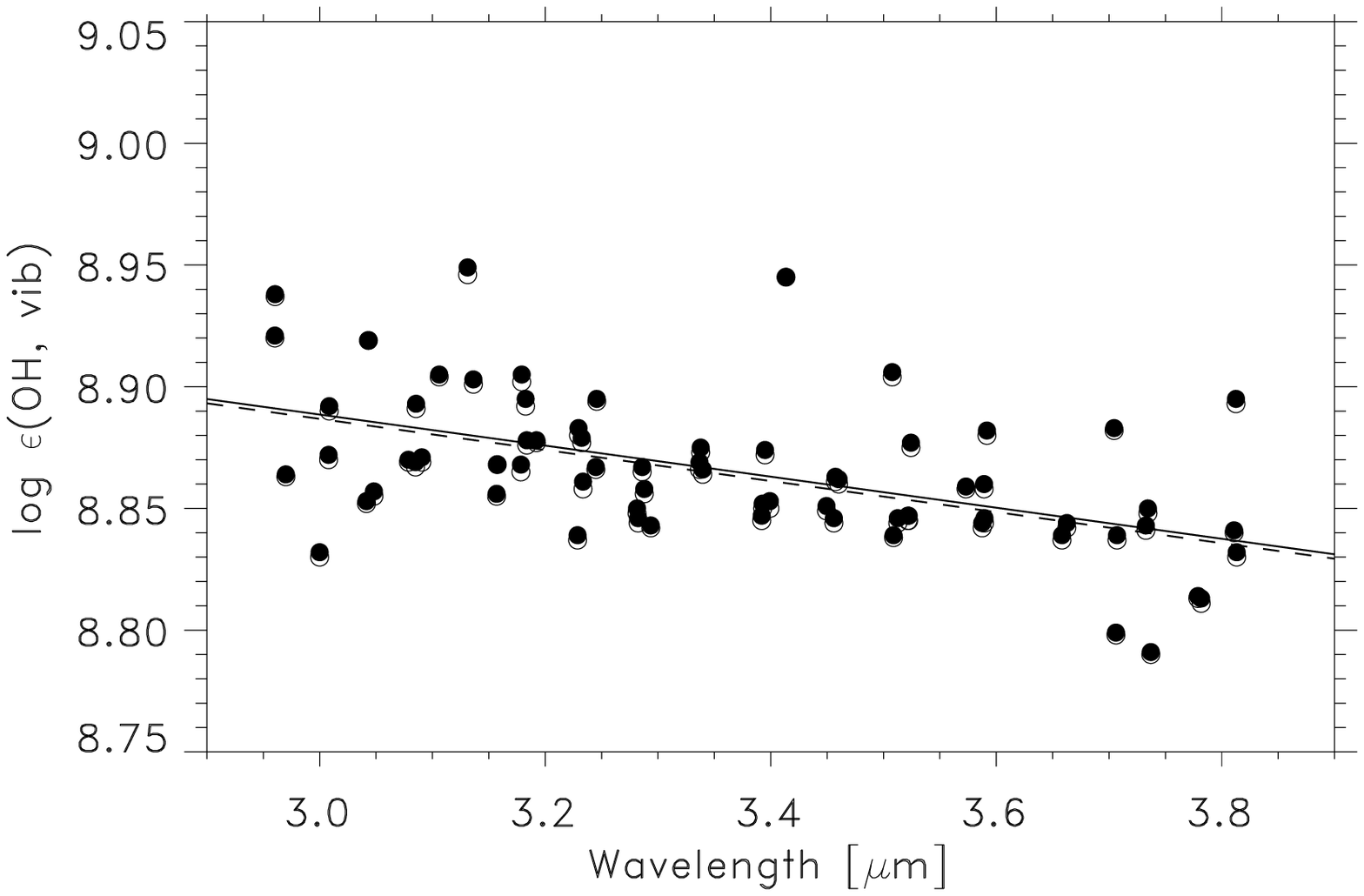}}
\resizebox{\hsize}{!}{\includegraphics{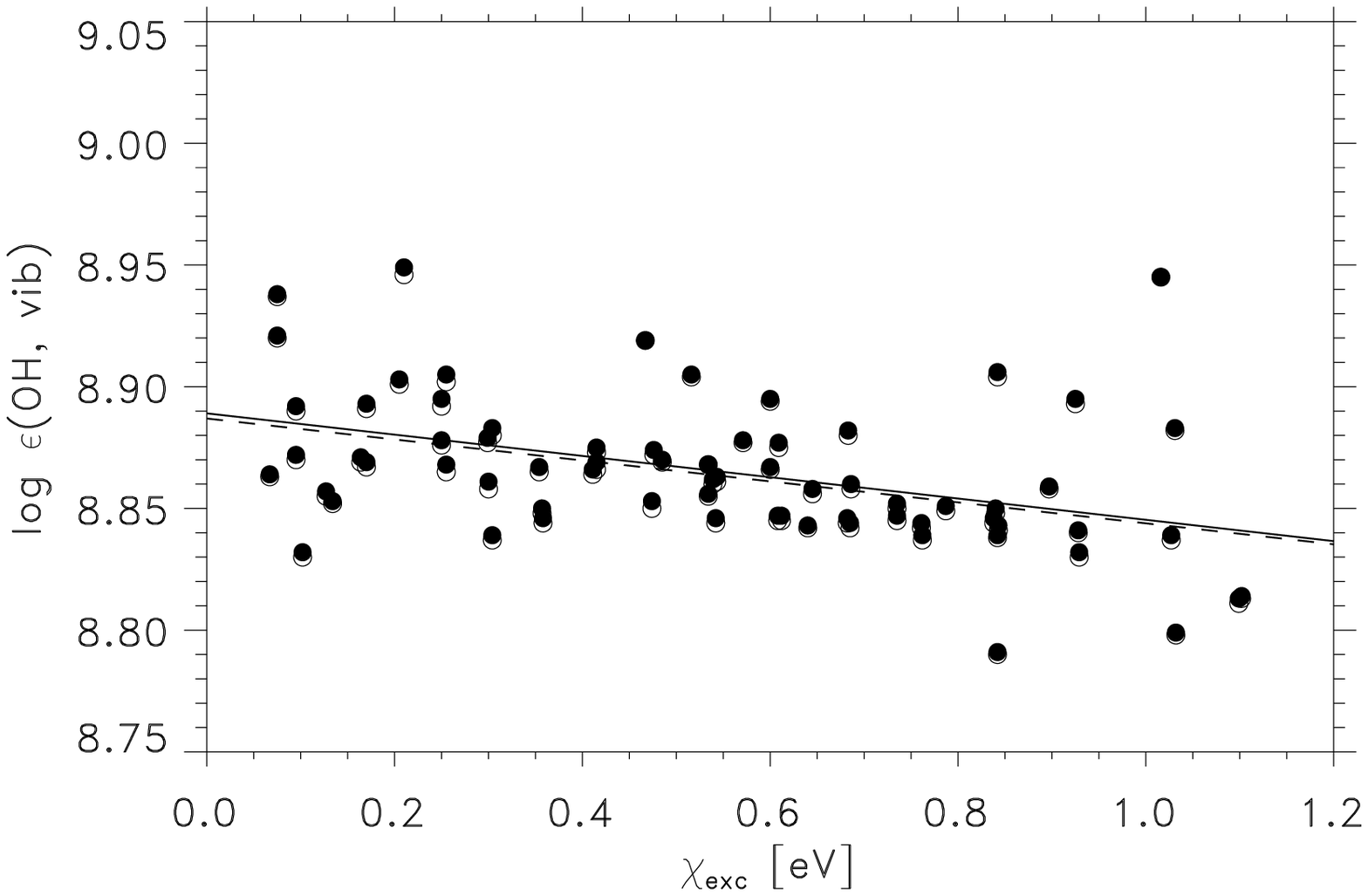}}
\resizebox{\hsize}{!}{\includegraphics{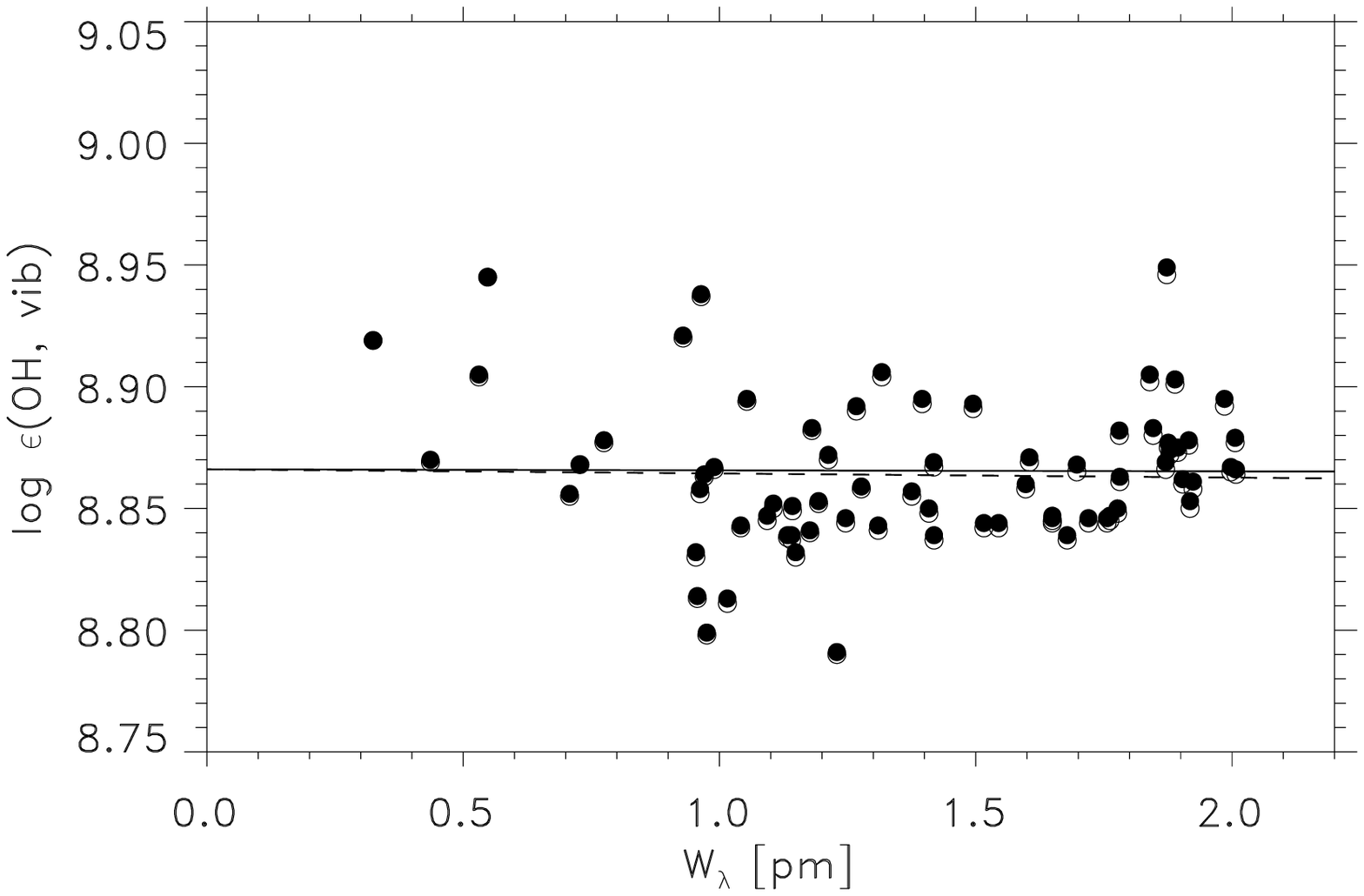}}
\caption{Same as Fig. \ref{f:OHvr_3D} but using the Holweger-M\"uller
(1974) semi-empirical model atmosphere and a microturbulence
of $\xi_{\rm turb} = 1.0$\,km\,s$^{-1}$ (filled circles with solid
lines denoting least-square-fits). The open circles
and dashed lines correspond to a larger microturbulence
($\xi_{\rm turb} = 1.5$\,km\,s$^{-1}$)
}
         \label{f:OHvr_hm}
\end{figure}

\begin{figure}[t]
\resizebox{\hsize}{!}{\includegraphics{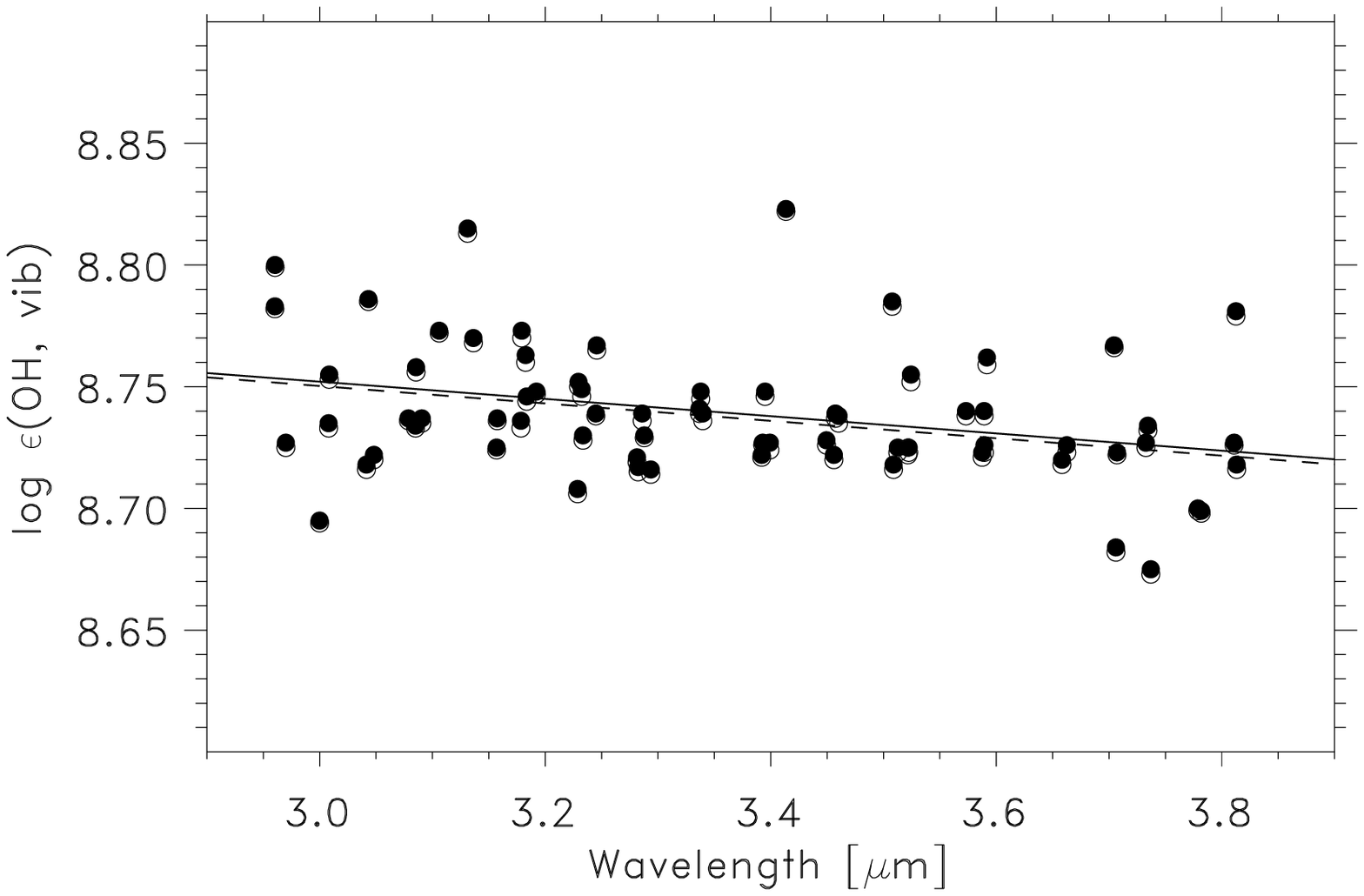}}
\resizebox{\hsize}{!}{\includegraphics{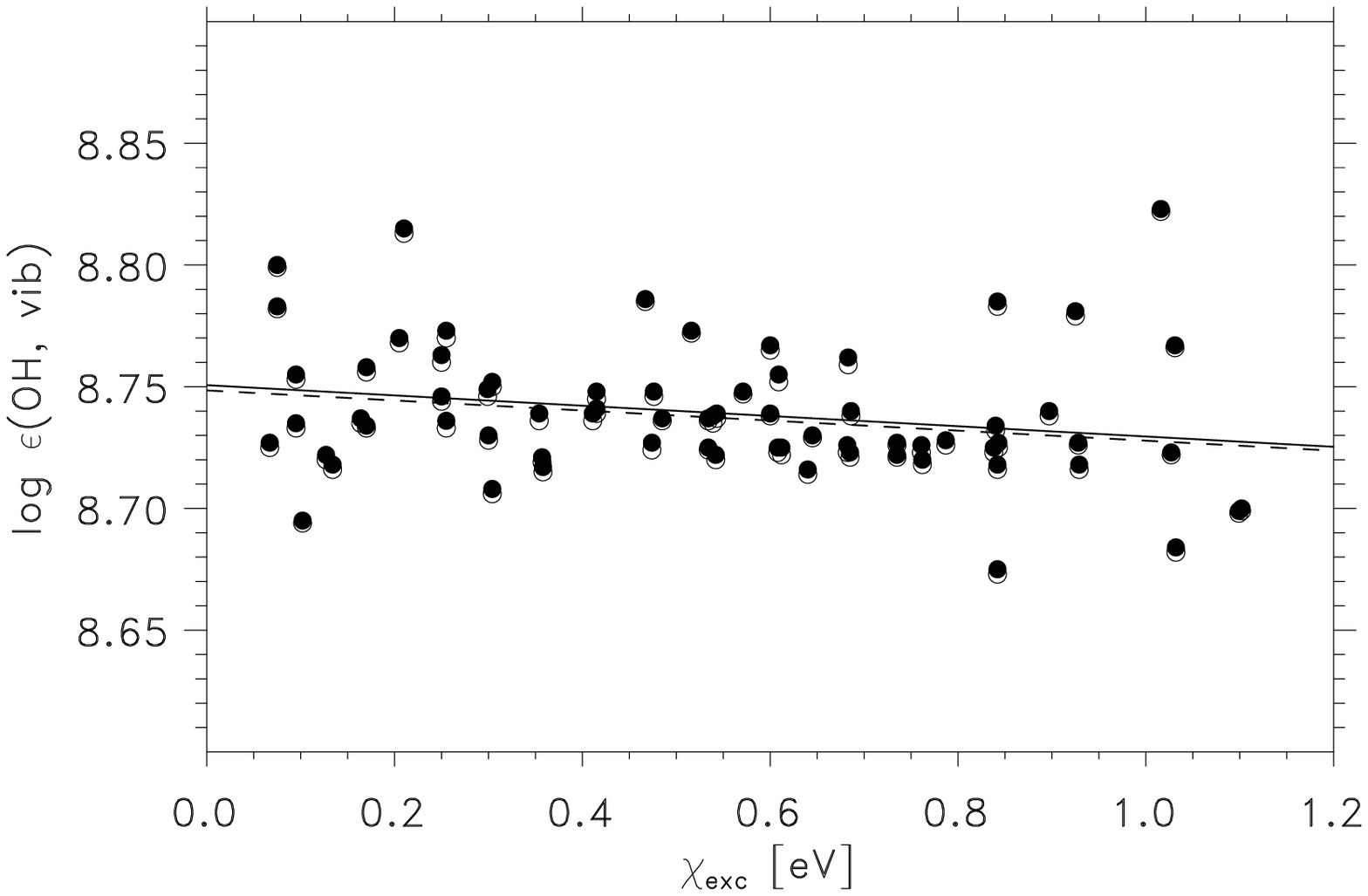}}
\resizebox{\hsize}{!}{\includegraphics{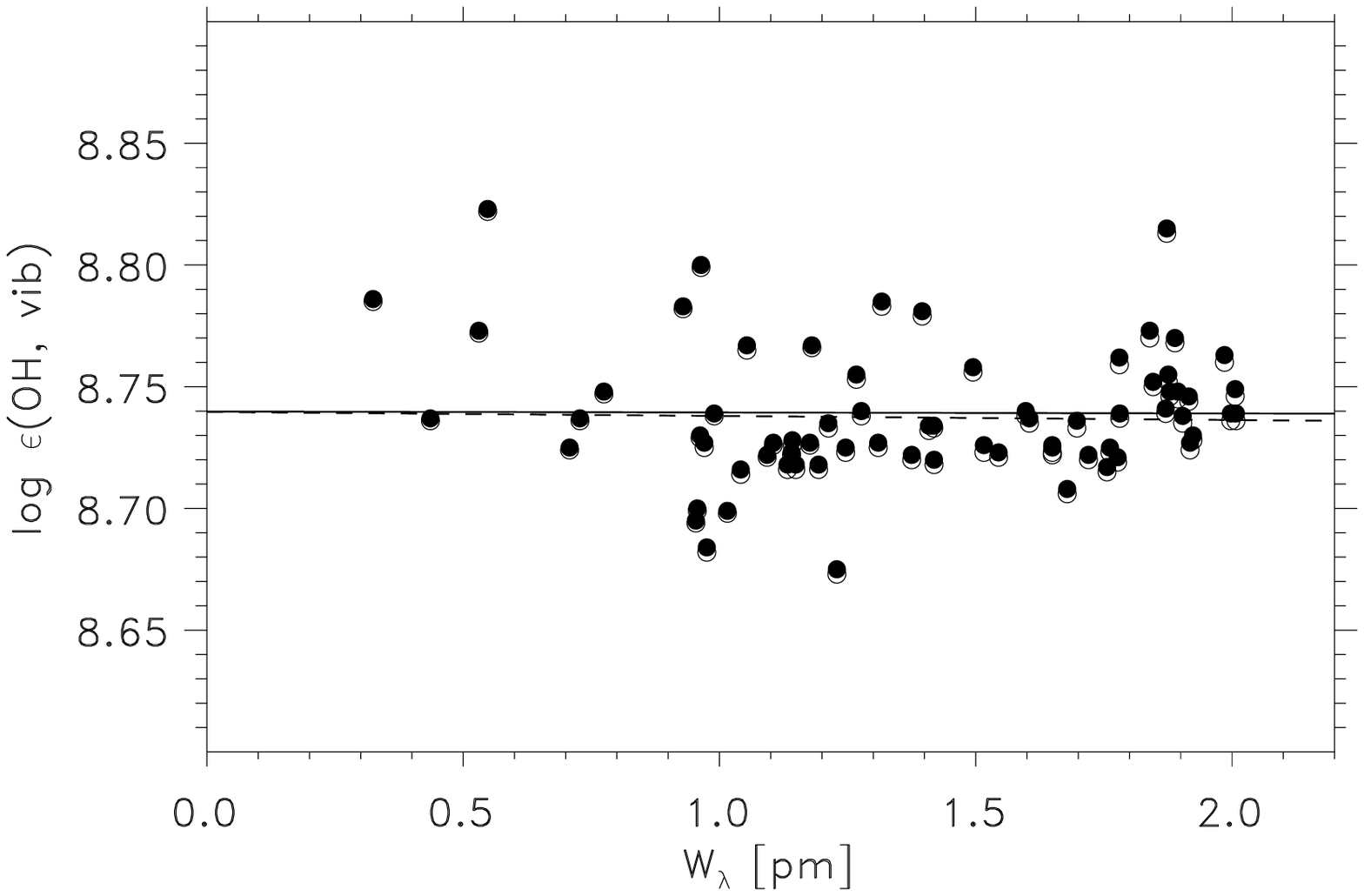}}
\caption{Same as Fig. \ref{f:OHvr_3D} but using 
a theoretical {\sc marcs} model atmosphere (Asplund et al. 1997) and a microturbulence
of $\xi_{\rm turb} = 1.0$\,km\,s$^{-1}$ (filled circles with solid
lines denoting least-square-fits). The open circles
and dashed lines correspond to a larger microturbulence
($\xi_{\rm turb} = 1.5$\,km\,s$^{-1}$)
}
         \label{f:OHvr_ma}
\end{figure}

The resulting O abundances with the 3D model atmosphere show
encouragingly small scatter and without any trends with neither wavelength,
excitation potential nor line strength, as clear from Fig. \ref{f:OHvr_3D}.
Considering the high degree of temperature sensitivity of OH molecule
formation and the minute fraction of oxygen tied up in OH, this excellent
agreement strongly suggests that the temperature
structure in the 3D model atmosphere is very close to reality in the
relevant atmospheric depth interval (typical mean optical depth of line
formation ${\rm log} \tau_{500} \approx -1.4$
in the case of the Holweger-M\"uller model. The concept of typical line
formation depth is of course less well-defined in the case of 3D models).
Giving equal weights to all 70 lines, the average of the individual
abundances yields log\,$\epsilon_{\rm O} = 8.61\pm 0.03$.

A similar analysis based on the 1D Holweger-M\"uller model atmosphere
gives a significantly larger abundance:
log\,$\epsilon_{\rm O} = 8.87\pm 0.03$. While the scatter is only
marginally larger than in 3D, there are disturbing trends in oxygen abundances
as function of wavelength and excitation potential
(Fig. \ref{f:OHvr_hm}); these trends are clearly inter-related as the low
excitation lines have shorter wavelengths.
It should also be noted that there is a significant trend with line strengths
when using equivalent widths measured in wavenumber units rather than wavelengths.
The relatively strong trend with excitation potential is particularly
revealing, implying that the temperature structure of the
Holweger-M\"uller model is inappropriate. In fact, the same conclusion
led Grevesse \& Sauval (1998) to experiment with a modified temperature
structure in order to remove the existing trends but thereby also
affecting the mean OH-based oxygen abundance.
In this respect, the 1D {\sc marcs} model atmosphere is performing better.
The resulting trends are barely significant (Fig. \ref{f:OHvr_ma})
and the scatter is very similar
to the 3D case. The {\sc marcs}-based abundance is
log\,$\epsilon_{\rm O} = 8.74\pm 0.03$.

\subsection{OH pure rotation lines}

The pure rotational lines of OH $X^2\Pi$ were first identified in the solar atlas
by Goldman et al. (1981, $v=0$), Blatherwick et al. (1982, $v=1$ and $v=2$)
and Sauval et al. (1984, $v=3$).
Goldman et al. (1983) and Sauval et al. (1984) have advocated for their
use as a prime abundance indicator of the solar oxygen abundance given the
large number of clean lines beyond $9 \mu$m. Our analysis utilises
127 of the weakest OH pure rotation lines belonging to the $v=0$, $v=1$, $v=2$,
and $v=3$ states with $N'' = 23-43$, whose equivalent widths have been measured from
the ATMOS solar IR atlas. Care has been exercised to only consider
OH lines minimally affected by telluric absorption.
While the equivalent widths and line depths for the pure rotation lines
are in general smaller than for the vibration-rotation lines, the pure
rotation lines are partly saturated since they are formed in significantly
higher atmospheric layers (typical mean optical depth of line
formation $-2.0\la {\rm log} \tau_{500} \la -1.6$
in the case of the Holweger-M\"uller model). 
Hence the pure rotation lines are sensitive
to the convective Doppler shifts (supposedly represented by microturbulence
in 1D analyses) in addition to the temperature sensitivity normal for all
molecules.

\begin{figure}[t]
\resizebox{\hsize}{!}{\includegraphics{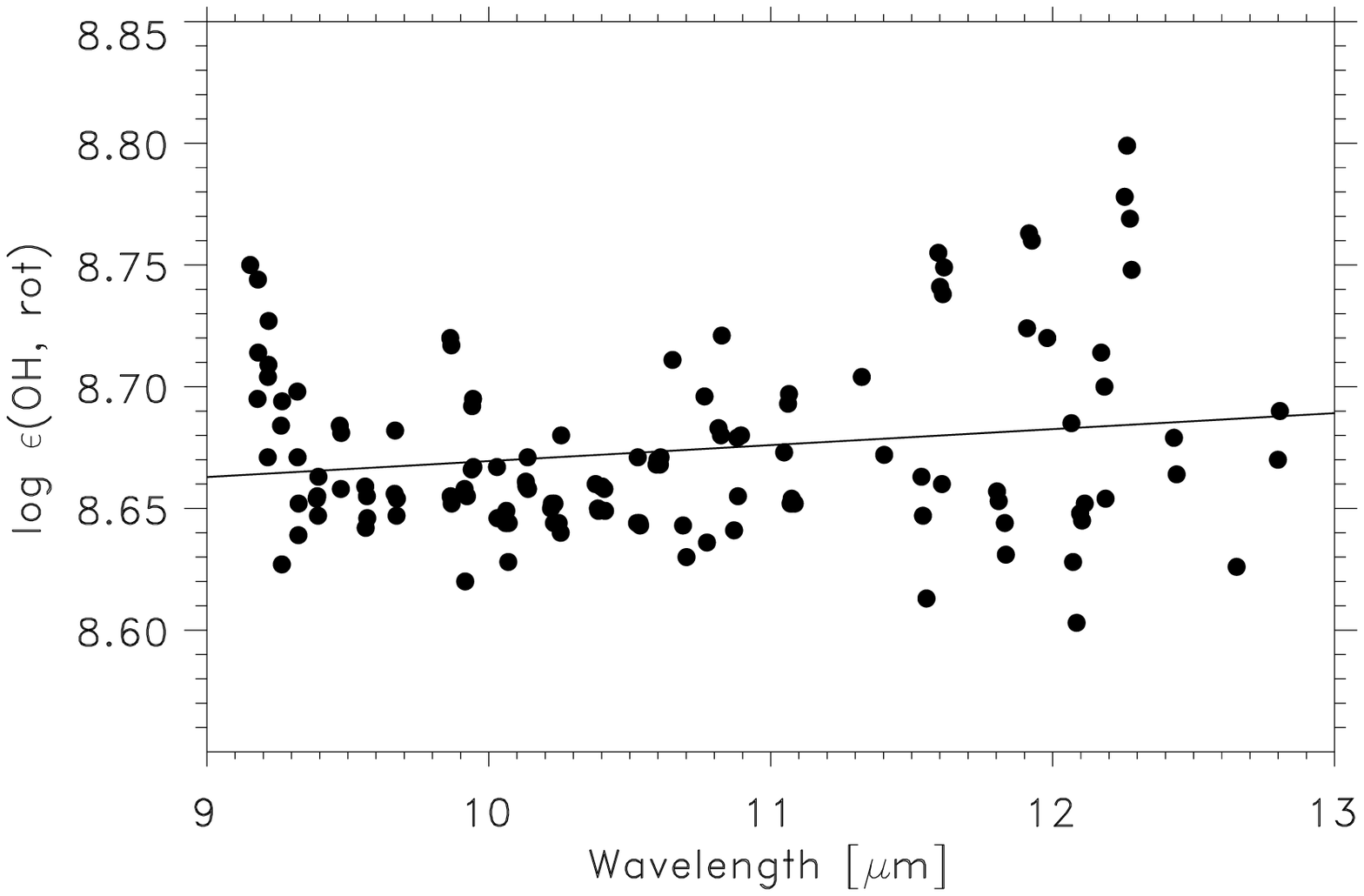}}
\resizebox{\hsize}{!}{\includegraphics{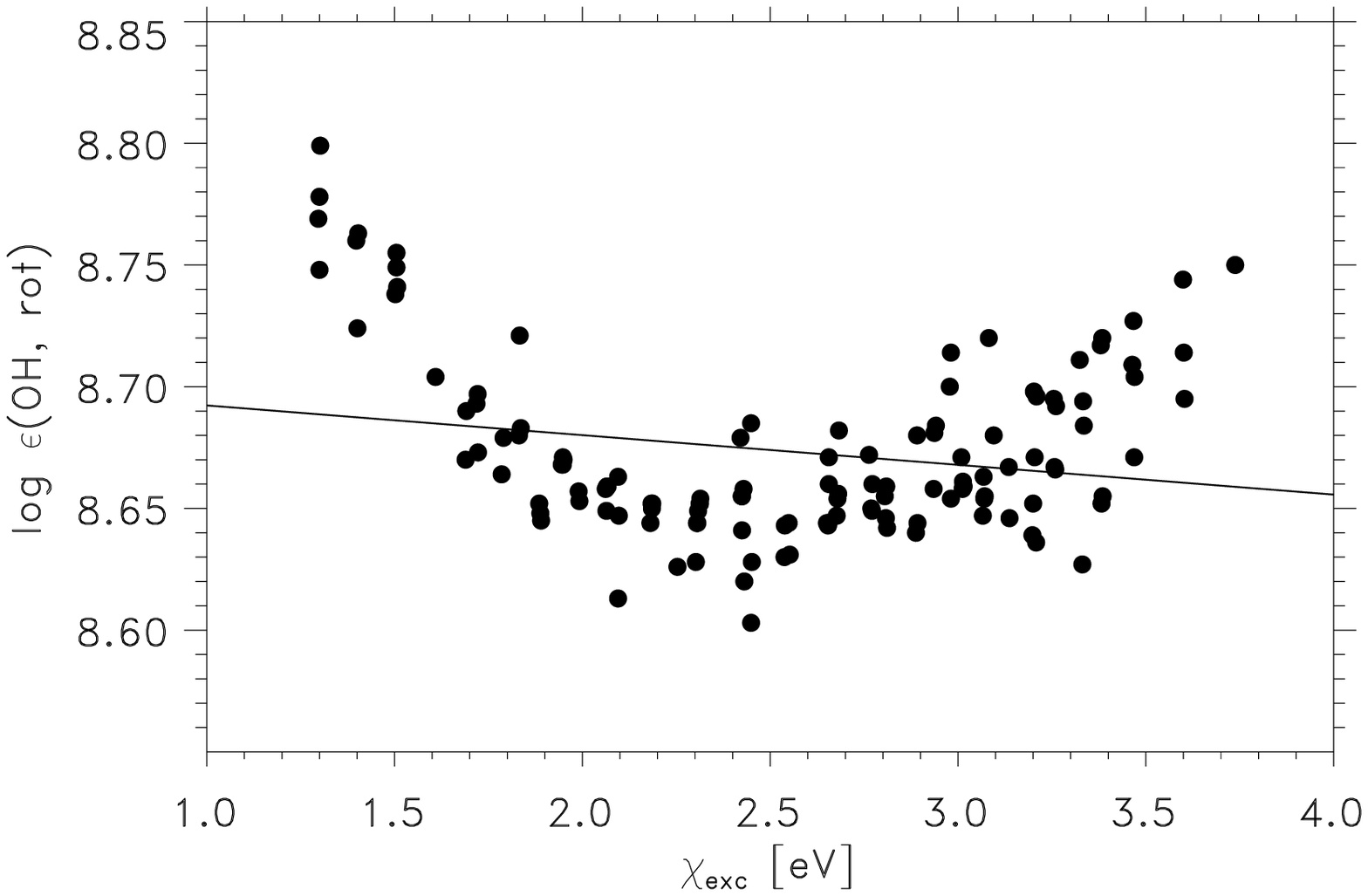}}
\resizebox{\hsize}{!}{\includegraphics{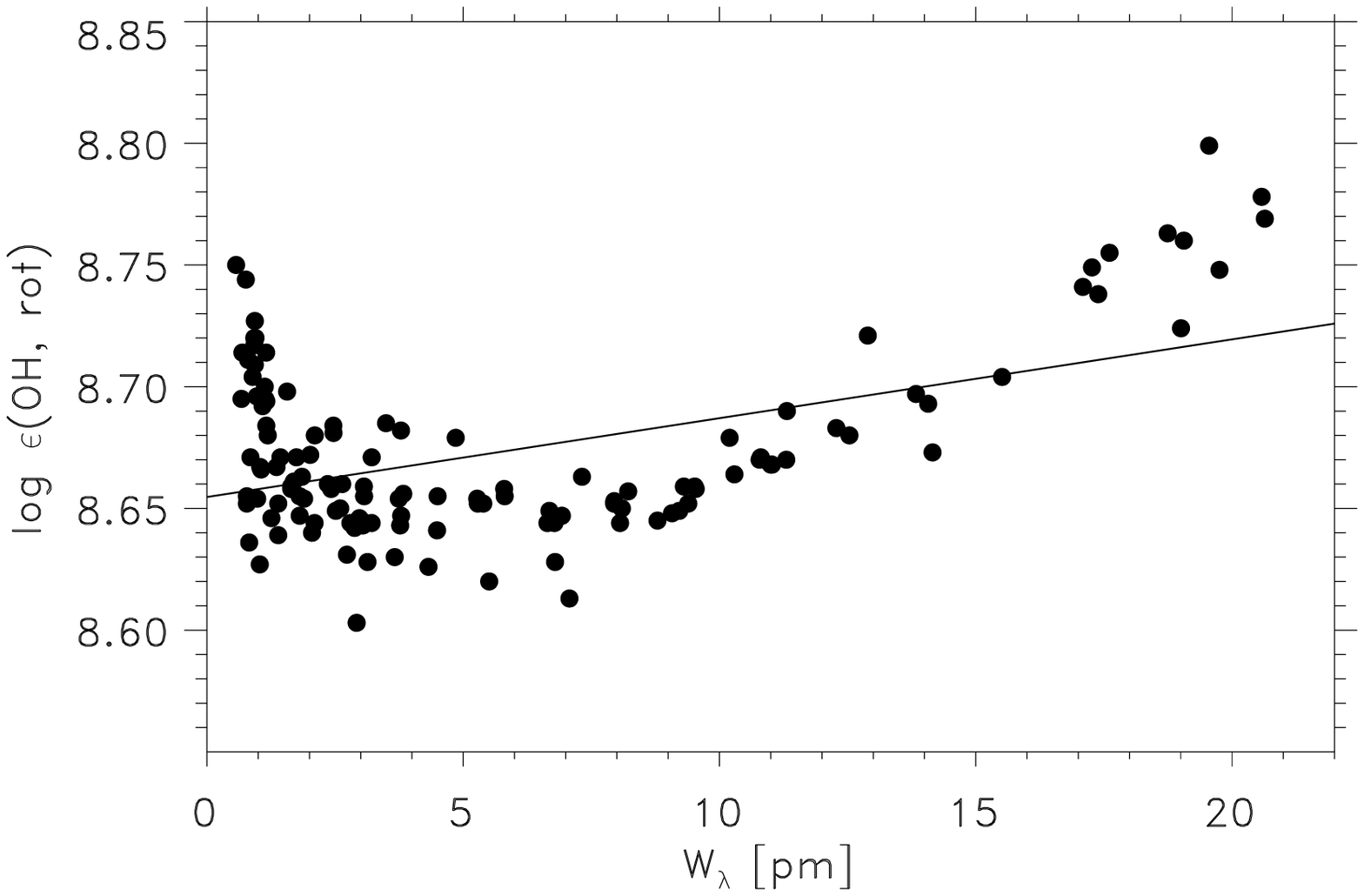}}
\caption{The derived solar oxygen abundance from OH pure rotational lines
using a 3D hydrodynamical solar model atmosphere as a function of
wavelength, lower level excitation
potential and line strength (in pm).
The solid lines denote the least-square-fits.
The rotational lines are more sensitive to the adopted microturbulence
than the vibration-rotation lines
since the stronger lines are partly saturated. The increased scatter for the
weakest lines (which correspond to the highest excitation lines)
is most likely due to increased observational difficulties in measuring
the equivalent widths accurately
}
         \label{f:OHrr_3D}
\end{figure}

The equivalent width-based O abundances derived with the 3D model atmosphere
show a small scatter, in particular for the lines with $W_\lambda = 1.5-10$\,pm.
Due to observational problems in accurately measuring the line strengths of
the weakest lines, the scatter increases significantly for $W_\lambda < 1.5$\,pm.
However, there are distinct trends with excitation potential
and line strength for lines with $\chi_{\rm exc} < 2$\,eV and
$W_\lambda > 10$\,pm, as seen in Fig. \ref{f:OHrr_3D}.
These two trends are directly related since the
lowest excitation lines are also the strongest.
It is expected that the realism of the 3D model 
atmosphere starts to deteriorate 
towards higher atmospheric layers, in particular due to the employed
radiative transfer treatment (total radiative heating/cooling 
determined by relatively few lines, lack of non-LTE effects,
no consideration of Doppler shifts for the energy balance etc).
These lines are also sensitive to the atmospheric velocity field.
Less weight should therefore be placed on the stronger rotational
lines. In the following,
the analysis is restricted to the pure rotation lines
with $W_\sigma = 1.5-10$\,pm. Giving equal weight to these
69 lines results in log\,$\epsilon_{\rm O} = 8.65\pm 0.02$.

\begin{figure}[t]
\resizebox{\hsize}{!}{\includegraphics{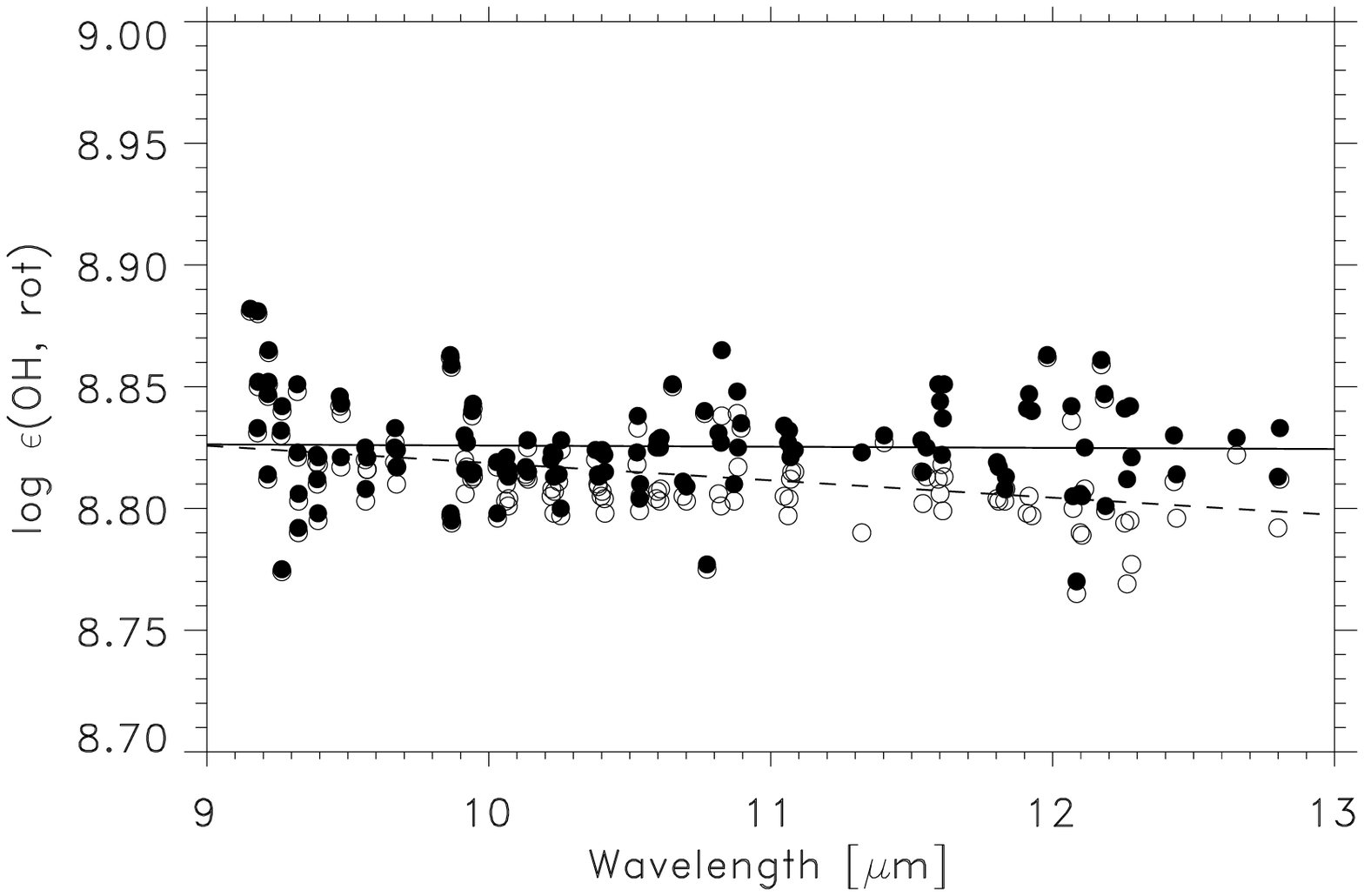}}
\resizebox{\hsize}{!}{\includegraphics{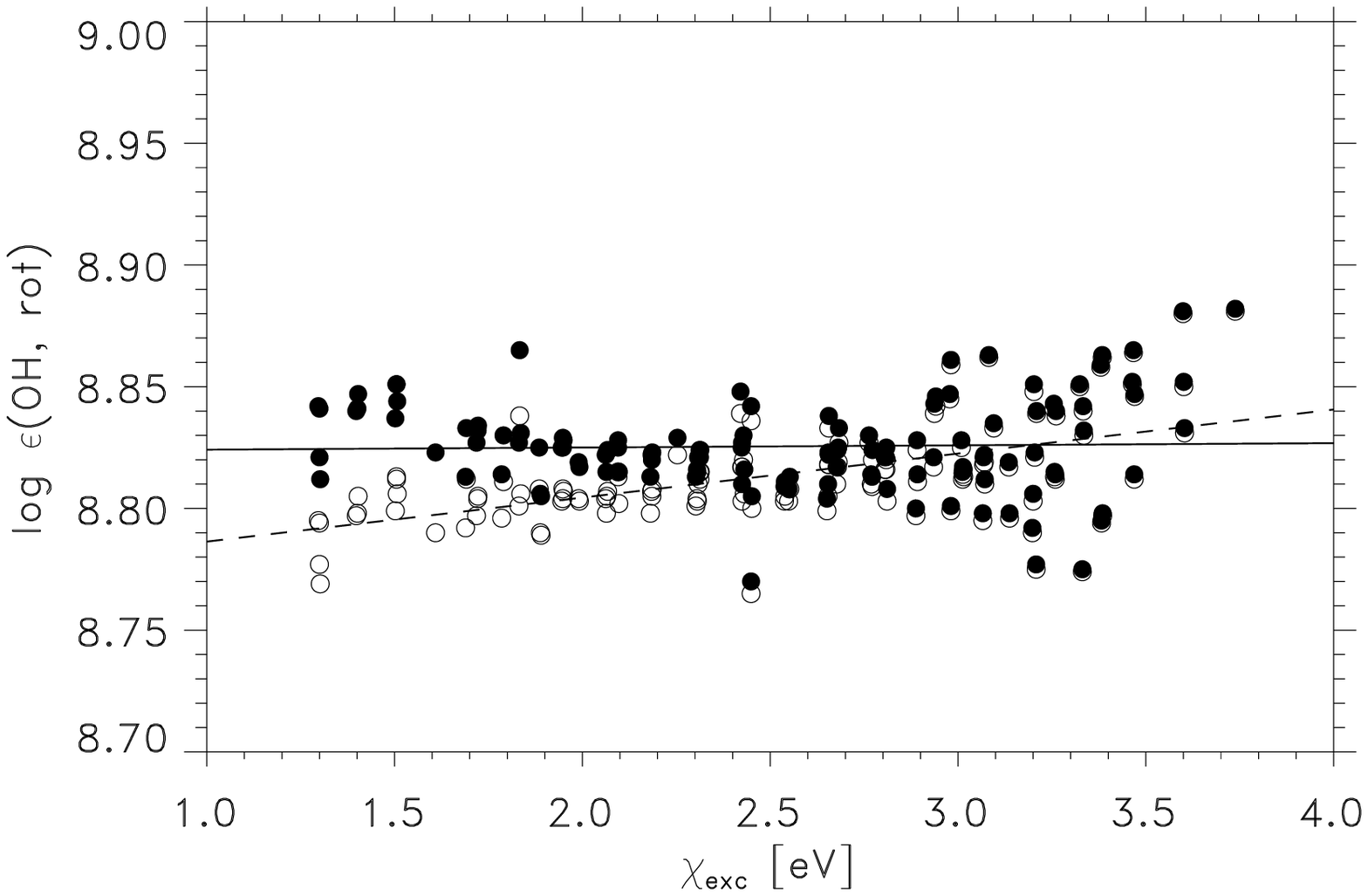}}
\resizebox{\hsize}{!}{\includegraphics{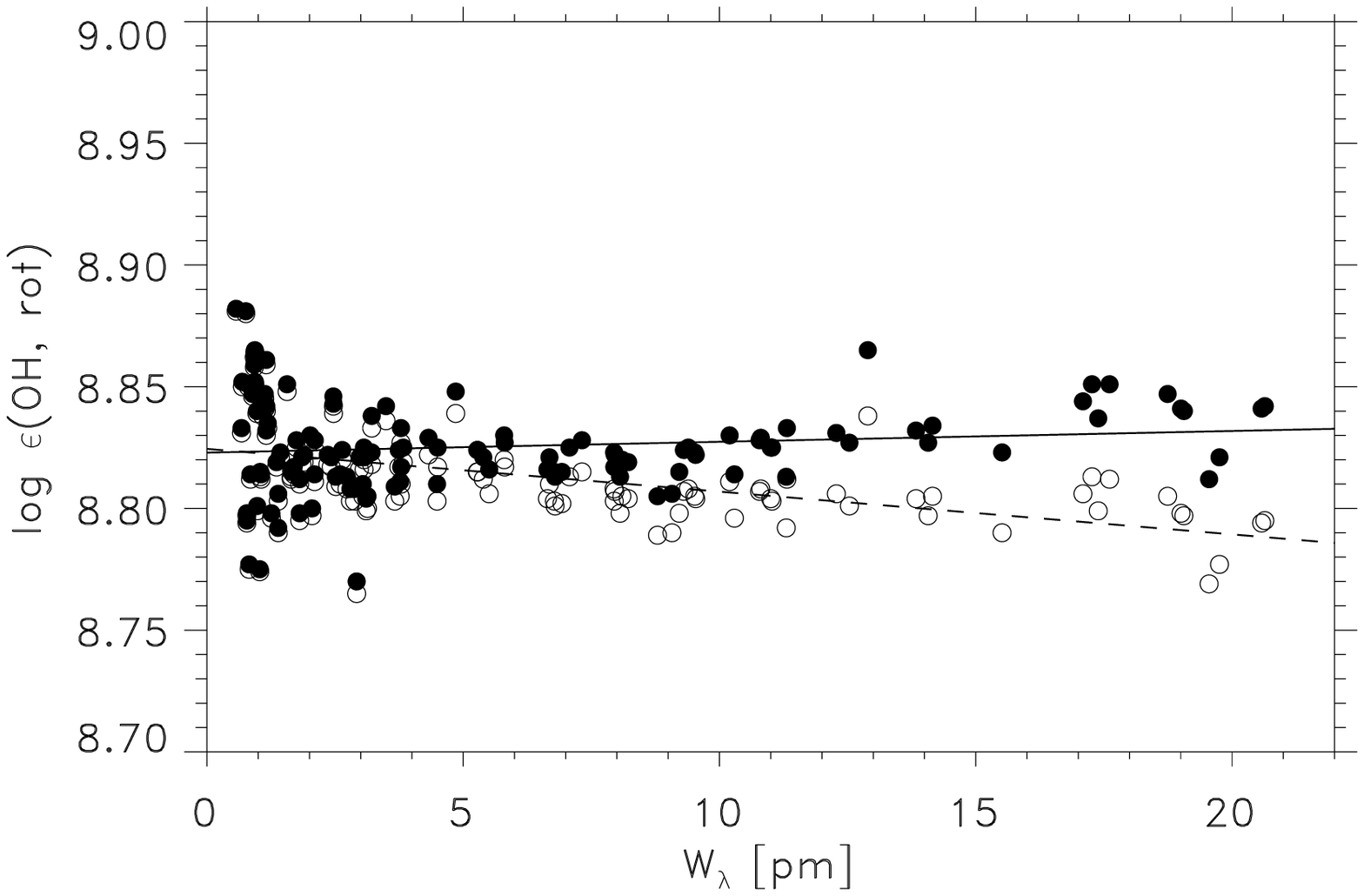}}
\caption{Same as Fig. \ref{f:OHrr_3D} but using the Holweger-M\"uller
model atmosphere. The filled circles correspond to
a microturbulence of $\xi_{\rm turb} = 1.0$\,km\,s$^{-1}$ while the
open circles are for $\xi_{\rm turb} = 1.5$\,km\,s$^{-1}$ with
the solid and dashed lines, respectively, denoting the least-square-fits
}
         \label{f:OHrr_hm}
\end{figure}

\begin{figure}[t]
\resizebox{\hsize}{!}{\includegraphics{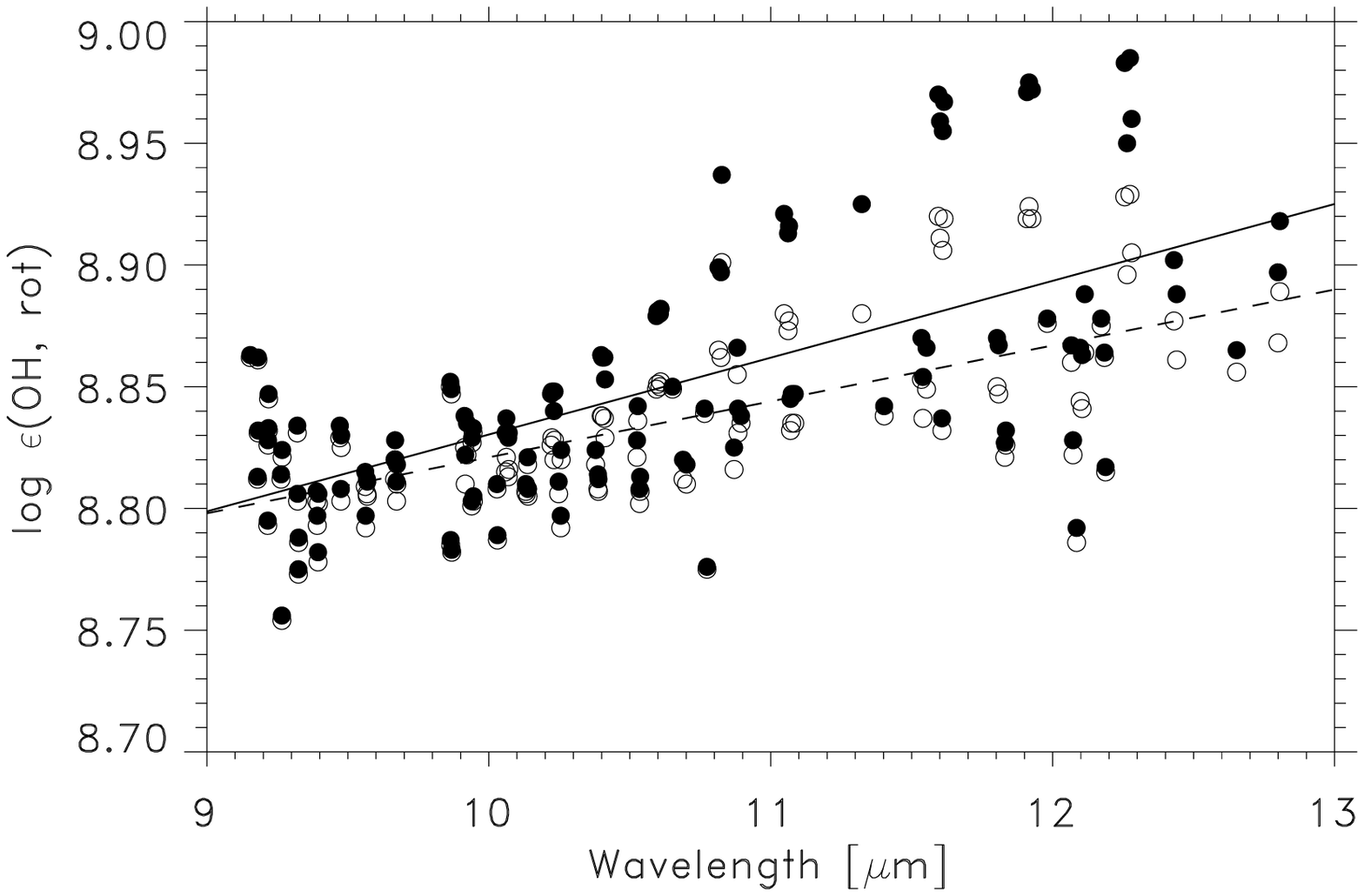}}
\resizebox{\hsize}{!}{\includegraphics{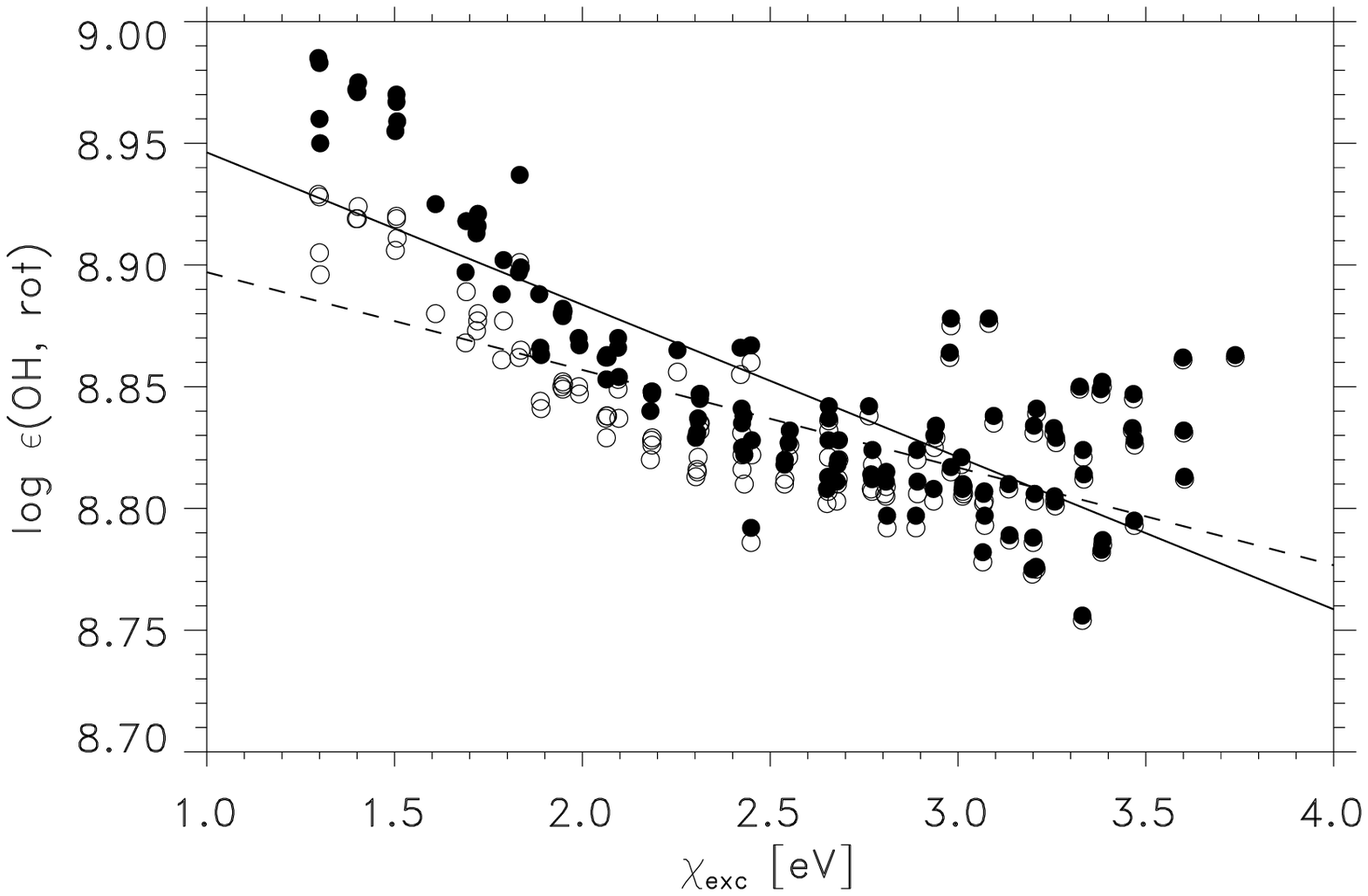}}
\resizebox{\hsize}{!}{\includegraphics{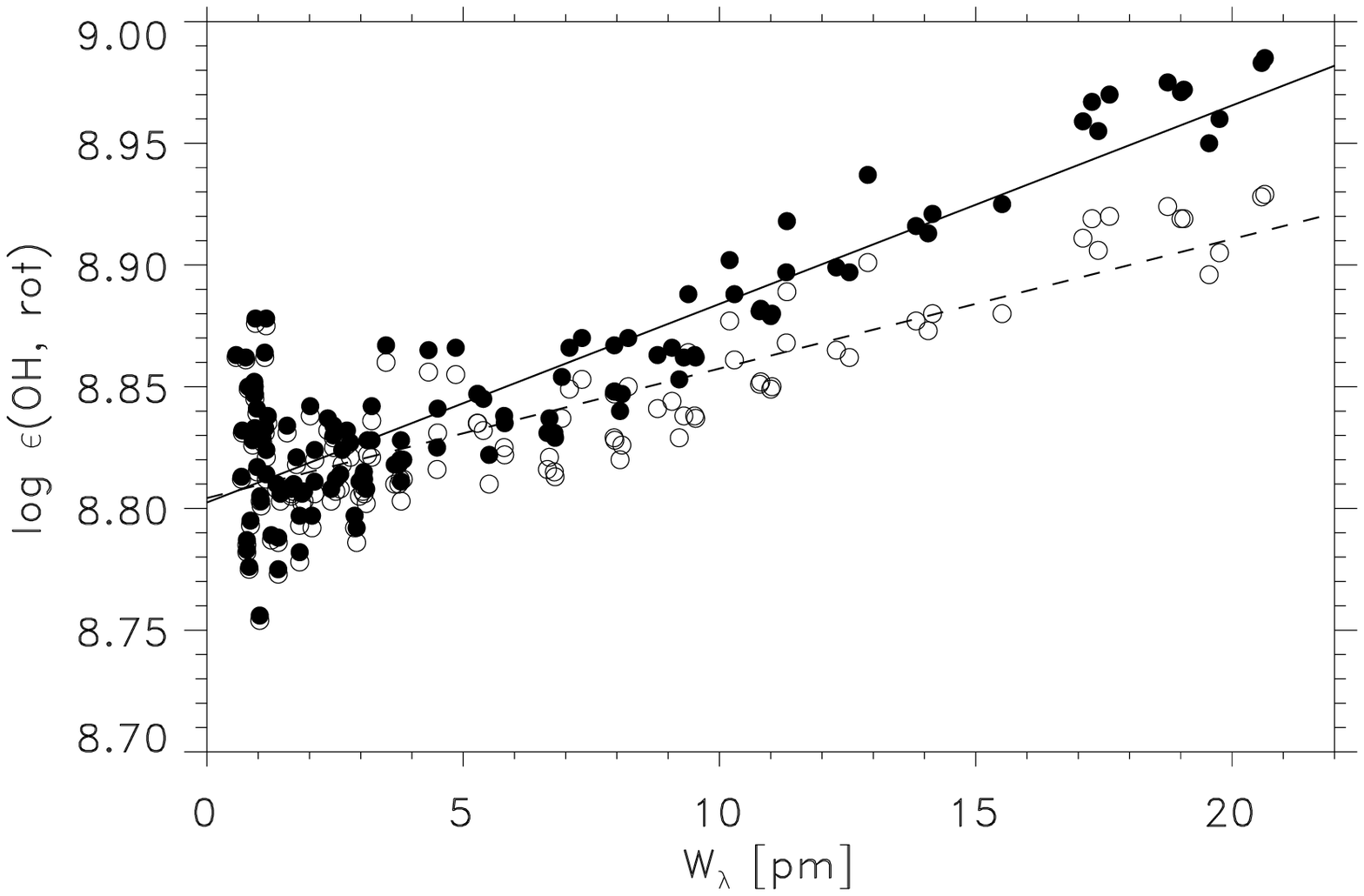}}
\caption{Same as Fig. \ref{f:OHrr_3D} but using 
a {\sc marcs} model atmosphere. The filled circles correspond to
a microturbulence of $\xi_{\rm turb} = 1.0$\,km\,s$^{-1}$ while the
open circles are for $\xi_{\rm turb} = 1.5$\,km\,s$^{-1}$ with
the solid and dashed lines, respectively, denoting the least-square-fits
}
         \label{f:OHrr_ma}
\end{figure}

We have also performed corresponding calculations with the 
{\sc marcs} and Holweger-M\"uller 1D model atmospheres 
(Figs. \ref{f:OHrr_ma} and \ref{f:OHrr_hm}).
The 1D analysis using the {\sc marcs} model atmosphere produces even
more spectacular trends with excitation potential and line strengths
with almost 0.2\,dex higher abundance for the strongest lines compared
with the weakest lines (Fig. \ref{f:OHrr_ma}).
Adopting instead a high microturbulence of
$\xi_{\rm turb} = 1.5$\,km\,s$^{-1}$ instead of the standard
$\xi_{\rm turb} = 1.0$\,km\,s$^{-1}$ only goes about half-way in
resolving the problems. Since such a high microturbulence is
inconsistent with similar analyses of other species, we conclude that
the temperature structure in these high atmospheric layers
in the {\sc marcs} model is clearly inappropriate for the analysis
of OH pure rotation lines.
The average abundance for the same 69 lines as for the 3D analysis is
log\,$\epsilon_{\rm O} = 8.83\pm 0.02$.
In this respect, the Holweger-M\"uller (1974) model is performing
remarkably well with essentially no trends with the standard
microturbulence of $\xi_{\rm turb} = 1.0$\,km\,s$^{-1}$
and a very small scatter (Fig. \ref{f:OHrr_hm}). This is somewhat surprising given
the problems encountered for the OH vibration-rotation lines
with the same model atmosphere (Sect. \ref{s:OHvr}).
However, the pure rotation lines are formed in higher layers than
the vibration-rotation lines (typically $-2.0 \la {\rm log} \tau_{500} \la -1.6$
and log\,$\tau_{500} \simeq -1.4$, respectively, in the case of the
Holweger-M\"uller model).
The average abundance with the Holweger-M\"uller model is
log\,$\epsilon_{\rm O} = 8.82\pm 0.01$.

As mentioned above, it is slightly surprising that the Holweger-M\"uller
model performs better than the 3D model for the pure rotation lines
while the opposite is true for the vibration lines.
The main effects must reside in the differences in temperature structure
and/or velocity field (convection in 3D and microturbulence in 1D).
The mean temperature
stratifications differ in the relevant atmospheric layers, as illustrated
in Fig. \ref{f:ttau}. It should be noted that the continuum 
optical depth scale at 3\,$\mu$m (typical wavelengths for vibration lines)
is very similar to that of 500\,nm,
while the continuum optical depth unity at 10\,$\mu$m (typical for rotation lines)
occurs at $\tau_{500} \approx -1.3$.
This also causes a reversal of the granulation pattern when viewed
at these long IR wavelengths: the bright (warm) material is located above
the intergranular lanes and the dark (cool) material lays above the
normal granules (see Fig. 1 in Kiselman \& Nordlund 1995).
Since the typical line formation depths differ for the two types of OH lines,
it is possible to remove the abundance trends present in Fig. \ref{f:OHrr_3D}
without tampering with the excellent consistency achieved in 
Fig. \ref{f:OHvr_3D}
by slightly decreasing the temperatures outside $\tau_{500} \approx -1.6$.
According to test calculations,
the necessary temperature modifications are relatively small and fully
within the uncertainties of the 3D temperature structure given the
inherent approximations for the energy balance in these high atmospheric 
layers. The main problem appears to be the relative fine-tuning of the
temperature alterations necessary in order not to perturb the vibration lines.
It is also possible that the resolution of the discrepancies can be
found in underestimated temperature inhomogeneities 
in the OH line forming region with the present 3D model. 
A detailed comparison of observed and predicted {\em absolute} disk-center
intensities is expected to be able to shed further light on this issue.
While the strongest rotation lines are sensitive to the velocity field,
the required changes make this a rather implausible solution according
to various test calculations.

\subsection{Summary}

As seen above and summarised in Table \ref{t:abund}, the different
oxygen diagnostics used for the present study yield very consistent
results when analysed with the 3D hydrodynamical solar model atmosphere.
Clearly the situation is much less satisfactory when relying
on classical 1D model atmospheres.
We see no compelling reason to prefer or exclude any of
the individual abundance indicators as the results are in all
cases quite convincing.
The quoted errors in Table \ref{t:abund} only reflect the
line-to-line scatter (standard deviation rather than standard deviation
of the mean as occassionally used for solar analyses) 
for the different types of lines. 
Using these standard deviations, the weighted mean becomes 
${\rm log} \, \epsilon_{\rm O} = 8.66 \pm 0.02$.
However, the total
error budget is most certainly dominated by systematic errors
rather than statistical.
Possible such sources include the atmosphere model (in particular
the temperature structure), the continuous opacity (due perhaps to
non-LTE effects on the H$^-$ molecule) and non-LTE effects on
the line and molecule formation.
To evaluate systematic errors is notoriously difficult.
An attempt of doing so can be achieved by a comparison of the
abundances implied by the different indicators, which in
the 3D case suggests that the remaining systematic errors
are under reasonable control given the very different types
of line formation processes involved and their widely different
temperature sensitivities.
A reasonable estimate would be to assign an uncertainty
of $\pm 0.05$\,dex on the derived solar oxygen abundance.
We therefore arrive at our best estimate of the solar oxygen
abundance as:
\nopagebreak
$${\rm log} \, \epsilon_{\rm O} = 8.66 \pm 0.05.$$

\begin{table}[t!]
\caption{The derived solar oxygen abundance as indicated by
forbidden [O\,{\sc i}], permitted O\,{\sc i} lines, OH vibration-rotation
line and OH pure rotation lines. The results
for the O\,{\sc i} lines include the non-LTE abundance corrections presented
in Table \ref{t:nlte}. The quoted uncertainties only reflect the line-to-line
scatter for the different types of O diagnostics
while no doubt the total error is dominated by systematic errors.
\label{t:abund}
}
\begin{tabular}{lccc}
 \hline
lines  & \multicolumn{3}{c}{${\rm log} \epsilon_{\rm O}$} \\
\cline{2-4}
         &   3D & HM & {\sc marcs} \\
 \hline
[O\,{\sc i}] & $8.68 \pm 0.01$ & $8.76 \pm 0.02$ & $8.72 \pm 0.01$ \\
O\,{\sc i}   & $8.64 \pm 0.02$ & $8.64 \pm 0.08$ & $8.72 \pm 0.03$ \\
OH vib-rot   & $8.61 \pm 0.03$ & $8.87 \pm 0.03$ & $8.74 \pm 0.03$ \\
OH rot       & $8.65 \pm 0.02$ & $8.82 \pm 0.01$ & $8.83 \pm 0.03$ \\
\hline
\end{tabular}
\end{table}

\section{The solar photospheric Ne and Ar abundances}

The solar abundances of neon and argon are also directly affected by the new solar
abundance of oxygen. Since spectral lines of Ne and Ar do not show up in the
solar photospheric spectrum, their solar abundances have to be deduced
from coronal matter. Oxygen, neon and argon particles are present in 
the solar wind as well as in solar energetic particles (SEP).
As they are high FIP (First Ionisation Potential) elements they are not
subject to fractionation affecting the low FIP elements
in coronal matter.
From the observed accurate SEP ratios, Ne/O = $0.152 \pm 0.004$ and
Ar/O = $(3.3 \pm 0.2) \cdot 10^{-3}$ (Reames 1999; these values agree with the 
somewhat less accurate solar wind data as well as with local galactic ratios derived
from hot stars and nebulae) we deduce revised values of the solar abundances
of Ne and Ar: 
log\,$\epsilon_{\rm Ne} = 7.84 \pm 0.06$ and
log\,$\epsilon_{\rm Ar} = 6.18 \pm 0.08$ respectively.
These values are again much smaller than the previously recommended values of
8.08 and 6.40, respectively, in Grevesse \& Sauval (1998).
The quoted errors account both for the measurement uncertainty in
the element ratios in SEP and the estimated systematic error in
the solar O abundance.

\section{Comparison with previous studies}

The derived solar oxygen abundances from the different indicators presented
in Sect. \ref{s:results} are particularly noteworthy for two reasons:
the significantly lower O abundance by as much as 0.25\,dex compared with
previously commonly adopted values (Anders \& Grevesse 1989; Grevesse \& Sauval 1998),
and the excellent agreement for all different features, in stark contrast
to what is achieved with 1D model atmospheres. In particular the latter fact
is a very strong argument in favour of the new generation of 3D hydrodynamical
model atmosphere and their application to accurate abundance analyses.

In spite of their weakness,
the forbidden [O\,{\sc i}] 630.0 and 636.3\,nm lines have often been advocated to be
the superior O abundance indicators, in particular the former line (Lambert 1978).
As both lines originate from the ground state, their line formations are
guaranteed to be exceedingly well approximated by LTE as verified by our non-LTE calculations.
Until recently, it was believed that the contribution from the blending Ni\,{\sc i}
630.0\,nm line was negligible, leading to an oxygen abundance of
log\,$\epsilon_{\rm O} \simeq 8.9$ with the Holweger-M\"uller semi-empirical
model atmosphere (Lambert 1978; Sauval et al. 1984; Holweger 2001).
Strong support for this high abundance has come from analyses of the OH
vibration-rotation (Grevesse et al. 1984) and pure rotation lines (Sauval et al. 1984),
which show an impressively small scatter between different OH lines
using the Holweger-M\"uller model. Together this led to the
conclusion that the solar O abundance is log\,$\epsilon_{\rm O} = 8.93 \pm 0.03$
(Anders \& Grevesse 1989). Since then the OH-based abundance has gradually
come down by 0.1\,dex in view of new preliminary analyses using a Holweger-M\"uller
model atmosphere with a slightly modified temperature stratification to
remove existing trends with excitation potential and line strength
(Grevesse \& Sauval 1998, 1999) and improved $gf$-values and measured
equivalent widths.
In the meantime, it has become clear that the
Ni\,{\sc i} blend for the [O\,{\sc i}] 630.0\,nm line may not be insignificant
after all.
Reetz (1998) made a very comprehensive investigation using both flux and
intensity solar atlases at different disk locations in an attempt to
constrain the Ni contribution. While no unanimous conclusion could be reached,
Reetz concluded that the Ni line could affect the results so much that the
solar O abundances may be as low as log\,$\epsilon_{\rm O} = 8.75$ when
analysed with the Holweger-M\"uller model.
A more precise estimate of the Ni line contribution was obtained by
Allende Prieto et al. (2001) by employing the same 3D hydrodynamical
solar model atmosphere as used here. In terms of abundance it was found
that properly taking the Ni line into account will lower the O abundance
by about 0.13\,dex. In comparison, the impact of moving from the classical
Holweger-M\"uller 1D model atmosphere to a 3D model was less important
in this case, 0.08\,dex, but the two effects compound each other by going
in the same direction. Together with a new $gf$-value (same as used here),
Allende Prieto et al. (2001) obtained log\,$\epsilon_{\rm O} = 8.69 \pm 0.05$.
The work presented here builds on the success of this study but extends
it greatly by also performing similar calculations for the other O
diagnostics. It is always worrisome when a result rests entirely on only
one spectral line, in particular when the implication in terms of
derived abundance is as profound as in this case and the line itself is
significantly blended.

While both the [O\,{\sc i}] and OH lines for long pointed towards a high solar
O abundance, the permitted high-excitation O\,{\sc i} lines were causing
severe problems, in particular the IR triplet at 777\,nm.
It is well-known that the triplet is susceptible to departures from LTE,
which increases the line strength and thus decreases the derived abundance.
The exact magnitude of this non-LTE effect depends on
the recipe, if any, for excitation by inelastic collisions with H.
The exact choice of model atmosphere and other details of the model atom
play a smaller role in this respect.
The essentials of the line formation process of the triplet lines are
in fact very well described even with a two-level approach
in as far as the main non-LTE effect is photon-losses in the line itself,
causing the line source function to deviate from the Planck value (Kiselman 1993).
A smaller contribution from a slight change in the line opacity also
modulates the result however.
Without accounting for H-collisions, the non-LTE abundance corrections
for these lines are about $-0.2$\,dex for the Sun, leading to a solar
O abundance of log\,$\epsilon_{\rm O} \simeq 8.7$ (Kiselman 1993).
Due to the large discrepancy in the non-LTE case with the [O\,{\sc i}] and OH
results, many authors have argued that the non-LTE calculations are
faulty. Indeed the LTE-abundance is close to those provided by the other diagnostics,
providing in some people's eyes a mean to {\em calibrate}
the poorly known H-collisions (Tomkin et al. 1992; Takeda 1994).
By necessity, such a procedure requires the H-collisions to be very large,
often even larger than the classical Drawin (1968) recipe for
such processes, thus enforcing efficient thermalization of the line and
a result close to the LTE-expectation.
It is important to realise, however, that the available laboratory and
theoretical calculations unambigously suggest that the Drawin recipe
over-estimates the H-collisions by about three orders of magnitude
in the sofar only studied cases of Li and Na
(Fleck et al. 1991; Belyaev et al. 1999; Barklem et al. 2003). 
If the same conclusions hold for O then 
the H-collisions are completely negligible in the solar case.
Furthermore, center-to-limb variations of the predicted LTE-profiles
(or those obtained from non-LTE calculations with Drawin-like H-collisions)
differ markedly from the observational evidence, as seen in
Fig. \ref{f:OI_limb} (see also Kiselman \& Nordlund 1995).

The root of the problem with 1D analyses has therefore been the poor
agreement between the O abundances estimated from the [O\,{\sc i}] and OH lines
on the one hand and the O\,{\sc i} lines on the other; with the Ni-induced downward
revision of the [O\,{\sc i}]-based abundance this statement must be modulated however.
In an important paper, Kiselman \& Nordlund (1995) attempted to resolve this
long-standing discrepancy through a pioneering 3D analysis of the
O\,{\sc i} triplet, the [O\,{\sc i}] 630.0\,nm line and a few OH vibration-rotation
and pure rotation lines.
They used a similar but less sophisticated 3D model atmosphere compared to
the one employed here and restricted the calculations to only two different snapshots.
A noteworthy finding of their study was that it may be possible to reconcile
the different abundance indicators in a proper 3D analysis since the
[O\,{\sc i}] and OH lines became stronger than in the
Holweger-M\"uller model atmosphere while the opposite was true for the triplet
as indicated by an equivalent two-level non-LTE calculation.
Due to the restricted numerical resolution and temporal coverage, they
were unable to perform an accurate abundance analyses for the lines, but
they concluded that
``{\em the solar oxygen abundance could very well be less than 8.80}''.
Our 3D model atmosphere covering some 100 snapshots over a 50\,min
solar-time sequence with an improved temperature structure and extension to
greater heights, allows us to derive significantly more accurate abundances
and finally settle the issue.

Recently, Holweger (2001) has independently re-visited the solar O abundance
using the [O\,{\sc i}] 630.0\,nm and O\,{\sc i} lines; he did not
study molecular lines. Holweger used the
old $gf$-value for the forbidden line and neglected the Ni-blend
and consequently found a high oxygen abundance:
log\,$\epsilon_{\rm O} = 8.92$ using the Holweger-M\"uller model. His seven
permitted lines suggested a significantly lower value, which, together
with non-LTE abundance corrections,
yielded log\,$\epsilon_{\rm O} = 8.73 \pm 0.04$.
The difference relative to our value with the Holweger-M\"uller model atmosphere
can be traced to differences in selected lines, adopted equivalent widths and much smaller
non-LTE corrections (for completeness
it should also be noted that Holweger uses intensity
rather than flux profiles).
In particular, Holweger adopts large H collision cross-sections, leading to small
non-LTE abundance corrections. As described above, the evidence is stacked against such
efficient H collisions.
In addition, Holweger added ``granulation corrections'' amounting to
$-0.02$\,dex on average to the derived abundance, finally arriving at
log\,$\epsilon_{\rm O} = 8.74 \pm 0.08$ when including
the [O\,{\sc i}] 630.0\,nm line.
It should be noted, however, that these granulation corrections
represent the difference in derived abundances between
full 2D spectrum synthesis calculations and corresponding 1D computations
using the 1D average of the 2D model atmosphere (Steffen \& Holweger 2002).
Thus, the abundance corrections are designed to 
only reflect the effect of temperature
inhomogeneities but not the difference in mean temperature structure
with for example the Holweger-M\"uller model atmosphere.
As a consequence, these granulation corrections can 
differ significantly both in magnitude and sign 
from those estimated by our method, which
also includes the effects of different mean temperature structures
between the 1D and 3D models. 
As our 3D solar model atmosphere has successfully passed a range of
observational tests (e.g. Stein \& Nordlund 1998: Rosenthal et al. 1999;
Asplund et al. 2000b; Asensio Ramos et al. 2003), we are confident 
that our approach is fully justified.

\section{Conclusions}

We have presented a determination of the solar oxygen abundance from
an analysis of [O\,{\sc i}], O\,{\sc i},
OH vibration-rotation and OH pure rotation lines by means of
a realistic time-dependent 3D, hydrodynamical model of the solar atmosphere.
All oxygen indicators yield very consistent results, in marked contrast with
studies employing classical 1D model atmospheres.
The here derived solar oxygen abundance of log\,$\epsilon_{\rm O} = 8.66 \pm 0.05$
(Table \ref{t:abund})
is significantly smaller than previously thought: the revision amounts to almost
a factor of two (0.27\,dex) compared with the often quoted value of
log\,$\epsilon_{\rm O} = 8.93$ from Anders \& Grevesse (1989),
and still much smaller than the value of 8.83 recommended more recently by
Grevesse \& Sauval (1998). 
This low abundance is strongly supported by the excellent agreement
between abundance indicators of very different temperature sensitivities
and line formation depths, and between
the observed and predicted line shapes and center-to-limb variations.
It should be noted that not the whole difference with previous results is attributed
to the employment of a 3D model atmosphere over classical 1D models since the
adoption of more recent $gf$-values, more realistic non-LTE procedures, better
observations and a proper accounting of blends also play an 
important role in this respect.

The new low solar oxygen abundance derived with the new generation of
3D hydrodynamical model atmospheres (e.g. Stein \& Nordlund 1998;
Asplund et al. 1999, 2000b; Asplund \& Garc\'{\i}a P{\'e}rez 2001)
resolves the problematic high metallicity of the Sun compared with
the solar neighborhood suggested by previous studies.
The oxygen abundance presented here (O/H\,$=460\pm50 \cdot 10^{-6}$)
is in very good agreement with that
measured in the local interstellar medium for realistic gas-to-dust ratios
(Meyer et al. 1998; Andr\'e et al. 2003): O/H\,$=410\pm60 \cdot 10^{-6}$.
In addition, the solar oxygen abundance is now similar to those
obtained from studies of nearby B stars 
(Cunha \& Lambert 1994; Kilian et al. 1994).
The most recent collation suggests that the mean B star value is 
O/H\,$=350\pm130 \cdot 10^{-6}$ (Sofia \& Meyer 2001)
but there is a significant scatter between
the existing B star analyses. 
Given the uncertainties we do not 
consider the remaining differences with our new solar oxygen abundance
as significant.
Finally, our value is close to those of nearby young, Galactic disk F and G dwarfs:
O/H\,$=450\pm160 \cdot 10^{-6}$ (Sofia \& Meyer 2001).
Interestingly, while the previous discrepancy most often has been blamed
on flaws in the analyses of hot stars and nebulae, the most serious shortcoming has
apparently rather been on the solar side.
We believe that 3D line formation calculations of late-type stars such
as those presented herein finally address this important issue, promising
much more accurate abundance analyses for the future.

There is, however, at least one area where the new solar O abundance together with
the corresponding downward revisions of the photospheric C, N and Ne abundances
(Asplund 2003, 2004; Asplund et al. 2004b,c)
does not resolve a problem but rather severely aggrevates it.
The accurate measurement of the solar oscillations, helioseismology,
allows a unique glimpse of the structure of the solar interior
(Christensen-Dalsgaard 2002). The analysis can be posed as an inversion
problem, enabling the extraction of the variation of the sound speed with
depth in the Sun. The predictions from standard models of the solar
structure and evolution show a remarkable agreement with the observed
sound speed from helioseismology to within
$\Delta c_{\rm s}/c_{\rm s} = 4 \cdot 10^{-3}$ (e.g. Christensen-Dalsgaard 2002).
The existing inversions are normally based on the opacities computed with
the old Anders \& Grevesse (1989) solar chemical composition,
or at least compositions with much larger values than our new results.  
With the new significant down-ward revisions of the solar C, N, O and Ne abundances,
the overall solar metallicity in the photosphere
decreases from $Z_{\rm old} = 0.0194$ (Anders \& Grevesse 1989) to $Z_{\rm new} = 0.0126$
(mass fraction of elements heavier than He).
This diminished metallicity propagates to a significant modification of the opacity
because C, N, O and Ne are main contributors to the opacity in different layers
below the convection zone,
and the sound speed is therefore directly affected.
As a result, the excellent agreement between the predicted and observed
sound speed variation with depth deteriorates markedly
(Pijpers 2002, private communication; see also Boothroyd \& Sackmann 2003).
We note that the new solar chemical composition should also affect the determination
of the solar He abundance in the outer convective zone 
from helioseismology inversion due to modifications of
the standard solar model. Likewise, the initial He abundance in the protosolar
cloud will likely be slightly smaller than the value $Y=0.27$ (mass fraction of He)
normally derived with existing solar evolution models based on the
old metal abundances.

At this stage it is unclear whether a resolution to this problem can be
found and if so where the solution may lay.
In the meantime, we offer the tentative suggestion that the problem
may be traced to the fact that the here obtained solar C, N and O abundances
only refer to the photospheric values rather than the abundances found
in the solar interior. If the diffusion of metals, as incorporated into
the stellar evolution models used for the solar inversion, is more efficient
than currently thought, it is conceivable that the good correspondence
in terms of the sound speed variation could be recovered.
The metallicity in the interior would then be significantly higher than
that of the convection zone and the photosphere.
The excellent concordance between all the different C, N and O diagnostics
achieved with the new generation of 3D hydrodynamical solar model atmospheres
in contrast with 1D analyses
certainly strongly argues for that the metallicity of the solar photosphere
is indeed significantly lower than previously thought.

\begin{acknowledgements}
MA has been supported by research grants from
the Swedish Natural Science Foundation, the Royal Swedish Academy
of Sciences, the G\"oran Gustafsson Foundation, and the Australian Research Council.
CAP gratefully acknowledges support from NSF (AST-0086321) and
NASA (ADP02-0032-0106 and LTSA02-0017-0093). 
NG is grateful to the Royal Observatory (Brussels) for financial support.
We wish to thank Mats Carlsson,
Remo Collet, Ana Elia Garc\'{\i}a P{\'e}rez, David Lambert, Arlette Noels,
\AA ke Nordlund, Frank Pijpers, Bob Stein and Regner Trampedach for
many helpful discussions and fruitful collaborations.
We also thank Joan Vandekerckhove for his help with the calculations at the Royal
Observatory (Brussels).
NSO/Kitt Peak FTS data used here were produced by NSF/NOAO. The GCT
was operated by the Universit\"ats/Sternwarte G\"ottingen at the Spanish
Observatorio del Teide of the Instituto de Astrof\'{\i}sica de Canarias.
We have made extensive use of NASA ADS Abstract Service.
\end{acknowledgements}

\end{document}